# Transformer-based cardiac substructure segmentation from contrast and non-contrast computed tomography for radiotherapy planning

Aneesh Rangnekar[a], Nikhil Mankuzhy[b], Jonas Willmann[a], Chloe Min Seo Choi[a], Abraham Wu[b], Maria Thor[a], Andreas Rimner[c], and Harini Veeraraghavan[a]

[a]Department of Medical Physics, Memorial Sloan Kettering Cancer Center, 1275 York Avenue, New York, NY 10065, USA

[b]Department of Radiation Oncology, Memorial Sloan Kettering Cancer Center, 1275 York Avenue, New York, NY 10065, USA

[c]Department of Radiation Oncology, Medical Center-University of Freiburg, Faculty of Medicine, University of Freiburg, German Cancer Consortium (DKTK), partner site DKTK-Freiburg, Freiburg, Germany

## Abstract

**Background and Purpose:** Accurate segmentation of cardiac substructures on computed tomography (CT) scans is essential for radiotherapy planning but typically requires large, annotated datasets and often generalizes poorly across imaging protocols and patient variations. This study evaluated whether pretrained transformers enabled data-efficient training using a fixed architecture with balanced curriculum learning.

**Materials and Methods:** A hybrid pretrained transformer-convolutional network (SMIT) was fine-tuned using lung cancer patients (Cohort I, N=180) imaged in supine position and validated on 60 held-out Cohort I patients and 65 breast cancer patients (Cohort II) imaged supine and prone. Two configurations were evaluated: SMIT-Balanced (32 contrast-enhanced CTs, 32 non-contrast CTs) and SMIT-Oracle (180 CTs). Performance was compared with nnU-Net and TotalSegmentator. Segmentation was assessed primarily using the 95th percentile Hausdorff distance (HD95) along with radiation dose and overlap-based metrics as secondary.

**Results:** SMIT-Balanced achieved comparable accuracy as SMIT-Oracle despite 64% fewer training scans; on Cohort I (HD95 of 6.6 $\pm$ 4.3 mm vs. 5.4 $\pm$ 2.6 mm), and Cohort II (10.0 $\pm$ 9.4 mm vs. 9.4 $\pm$ 9.8 mm), and was robust to patient, imaging, and data variations. Radiation dose metrics derived from SMIT were equivalent to manual delineations. nnU-Net improved over the publicly trained TotalSegmentator but showed reduced cross-domain robustness compared to SMIT.

**Conclusions:** Balanced curriculum training reduced labeled data requirements without compromising accuracy relative to the oracle model and maintained robustness to patient and imaging variations. Pretraining reduced dependence on data domain and obviated need for data-specific reconfiguration architecture as required by nnU-Net.

### Graphical Abstract

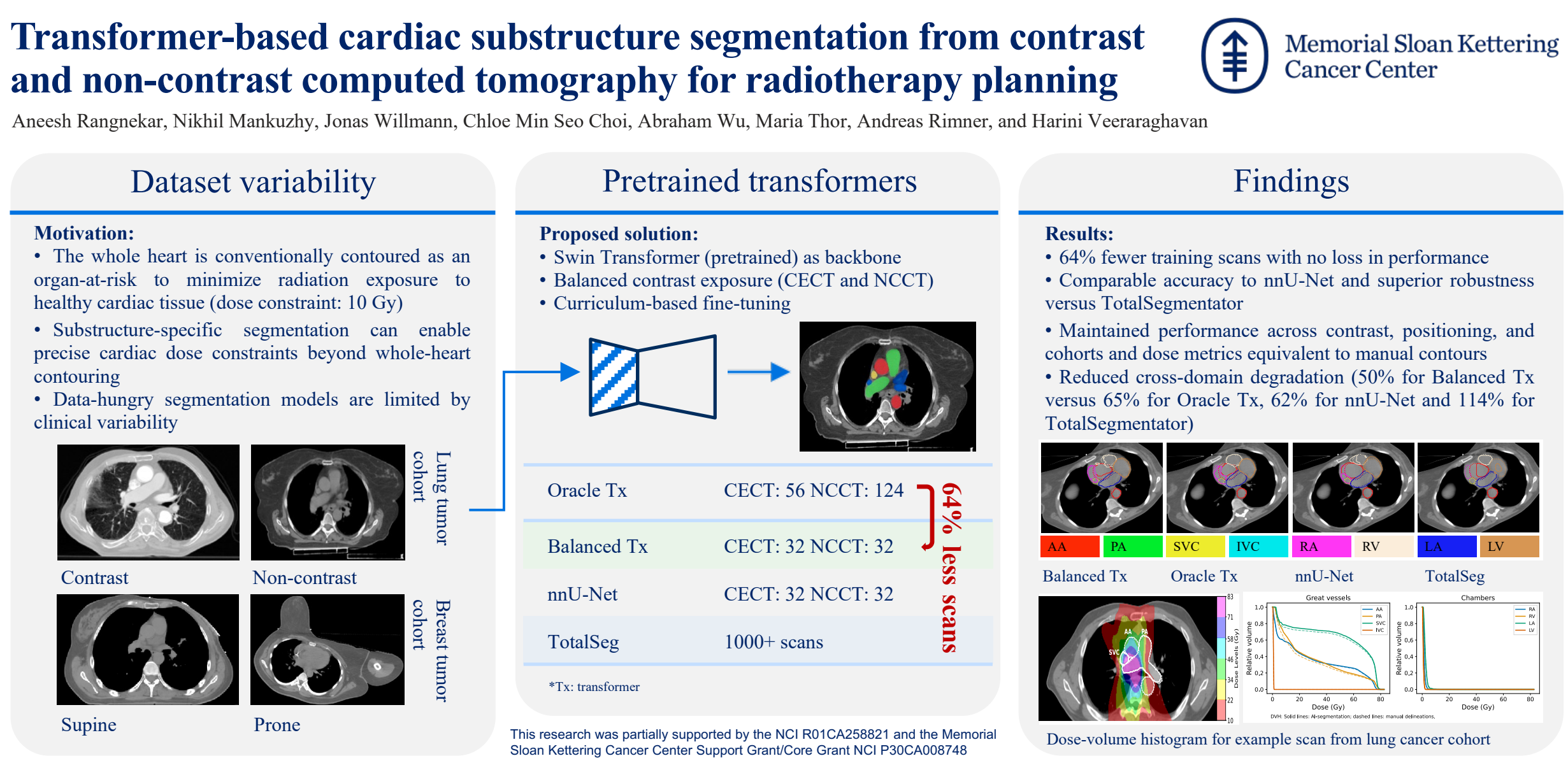

## 1. Introduction

Radiation exposure to cardiac substructures from radiation treatment (RT) is associated with the development of acute and late cardiac toxicities and subsequent shorter survival in patients with lung [1-3], esophageal, and breast cancers [4]. Current treatment planning approaches typically consider the heart as a single organ, which is insufficient to differentiate the radiation sensitivities of individual cardiac substructures [5]. Availability of accurate automated segmentation methods could reduce manual delineation requirements and allow for substructure-specific dose constraints.

Models that can produce robustly accurate segmentations for common clinical variations such as CT contrast, patient anatomy, and scan positions are essential for providing uninterrupted and consistently accurate automated radiotherapy plans. Data-specific automated network reconfiguration introduced in nnU-Net [6] that dynamically modifies required number of network layers and preprocessing based on training data has demonstrated capability to provide highly accurate segmentation of thoracic organs [7-8] and cardiac substructures [9-14] from both computed tomography (CT) and magnetic resonance images. nnU-Net is particularly attractive owing to ease of use in a turn-key fashion and obviates the need to build specialized architectures to enhance accuracy on testing data that is in-distribution with training data. However, data-specific preprocessing may reduce generalization under drifts in the data domain. Moreover, radiation treatment planning involves varying numbers of organs depending on treatment modality, which would invariably result in task-specific architectures, potentially complicating AI quality assurance and maintenance. A fixed architecture with transfer learning capabilities with data domain invariant performance would simplify clinical deployment while maintaining acceptable accuracy.

Recent advances in self-supervised learning (SSL) pretraining of vision transformers demonstrated that pretrained representations can be fine-tuned with fewer labeled data while maintaining fixed architectures [15-17]. However, whether fixed-architecture approaches can achieve clinically acceptable accuracy compared to self-reconfiguring methods remains unclear, particularly for cardiac substructures, which exhibit anatomic and appearance variability due to imaging differences.

This work tested the hypothesis that a hybrid transformer convolutional network (SMIT) leveraging SSL pretrained encoder can enable data-efficient and domain invariant cardiac substructure segmentation. We tested the hypothesis by measuring: (a) accuracy with different numbers of training data, (b) robustness to common imaging (contrast versus non-contrast; prone versus supine) and patient (male versus female; breast versus lung cancer) variations as well as data domain differences (institutional training and testing versus public data training and institutional data testing), and (c) benchmarking performance against an oracle model trained with all available training data and self-reconfiguring nnU-Net.

## 2. Materials and Methods

### 2. 1. Patient cohorts

Two retrospectively collected institutional patient cohorts were analyzed following approval of the local Institutional Review Board. Cohort I, used for network training and validation, consisted of 240 CT scans from patients with locally advanced non-small cell lung cancer treated with conventionally fractionated radiation treatment. Patients received 6 MV photons using five-field intensity-modulated radiation therapy (IMRT), typically delivered over 30 fractions at 1.8-2.0 Gy per fraction between 2004 – 2014 [18-20]. Treatments were planned with an in-house treatment planning system, with dose recalculated using the Eclipse TPS (Varian Medical Systems, Palo Alto, CA) [21] with inhomogeneity correction applied. All patients were positioned head-first supine, and one 4DCT and one free-breathing CT were acquired at the time of treatment planning. The whole heart volume was dose-constrained during optimization. Cohort I was split into training (N=180) and validation (N=60) sets, stratified by use of contrast agent.

Cohort II, used as an external test set, comprised 65 patients with stage I-III breast cancer who received radiation treatment [22]. Patients were treated with either 3D conformal RT (N=58) or static/rotational IMRT (N=7), typically at 2.65 Gy over 16 fractions between 2007 – 2021. Treatments were planned with an in-house TPS prior to 2014 and Eclipse TPS (v15.0; Varian Medical Systems) thereafter. Deep-inspiration breath hold was used for left-sided tumors and typically also for right-sided tumors. The whole heart was dose-constrained during optimization.

In addition, publicly available CT data from the TotalSegmentator dataset [23] consisting of 1,204 CT scans with publicly provided segmentations were used to train SMIT and nnU-Net for assessing domain generalization.

### 2. 2. Imaging characteristics

Cohort I consisted of a mix of contrast-enhanced CT (CECT – N=80) and non-contrast CT (NCCT – N=160) scans of varying image quality and resolutions across different GE and Siemens scanners. Image acquisition used a kilovoltage peak range of 120 – 140, with an average voxel spacing of 1 mm x 1 mm x 3 mm. All patients were scanned in the supine position. Cohort II consisted of only NCCT scans, with patients scanned in either supine (N=45) or prone (N=20) positions. The scans from TotalSegmentator dataset [23] were natively resampled to 1.5 mm isotropic resolution.

## 2. 3. Delineation guidelines

Eight cardiac substructures were manually delineated on all scans from Cohort I and II: great vessels (aorta, pulmonary artery, superior vena cava, inferior vena cava) and cardiac chambers (right atrium, right ventricle, left atrium, left ventricle). Contouring followed institutional guidelines based on Feng et al. [24], and all delineations were verified and refined as necessary by an expert radiation oncologist for quality assurance.

## 2. 4. Deep learning method

Figure S1 illustrates the hybrid transformer-convolutional SMIT network architecture. The transformer backbone used publicly available Swin Transformer checkpoint [15,16] (Supplementary Material A). The convolutional U-Net decoder [25] was randomly initialized with skip connections linking transformer backbone features for spatial detail preservation.

Two primary training configurations were evaluated: a data-efficient 'SMIT-Balanced' using 32 CECT and 32 NCCT scans, and fully supervised 'SMIT-Oracle' using all 180 training scans to estimate the performance ceiling (upper bound). To analyze the contribution of training design choices, we performed a series of ablation experiments derived from the data-efficient configuration: (1) freezing the encoder during fine-tuning, (2) training with contrast-only data. (N=56), and (3) training with non-contrast-only data, with NCCT cohort size (N=56) matched to CECT availability. The CECT/NCCT models were limited by the number of scans available; the small difference in total training size relative to the balanced configuration (N = 64) falls within the performance plateau identified in Figure S2 (< 0.3 mm HD95 gain between 56 and 64 scans). Finally, to evaluate cross-data generalization, SMIT (without any modifications) was trained on the external dataset from TotalSegmentator and tested on both study cohorts.

All models were trained on random $128^3$-voxel crops using equally weighted soft Dice and cross-entropy loss, optimized with Adam over 32,000 iterations (Supplementary Material B). Inference was performed on full-resolution scans using 3D sliding-window evaluation ($128^3$ crops with 50% overlap) on held-out test sets of 60 Cohort I scans and all 65 Cohort II scans. Segmentation experiments were implemented in MONAI (v0.8.1) [26] and PyTorch (v2.0.1) [27].

## 2. 5. nnU-Net comparison models

SMIT-Balanced was also benchmarked against nnU-Net [6] trained using the exact same data, with same training/validation splits and evaluation metrics. nnU-Net followed its standard automatic configuration protocol with architecture and preprocessing, resulting in an input size of 192 x 192 x 64 voxels at 1 x 1 x 3 $mm^3$ spacing, and was trained for 250,000 iterations (Supplementary Material C).

We additionally evaluated the publicly available nnU-Net–based TotalSegmentator model [23]. The *heartchambers_highres* variant [28] was used for the aorta, pulmonary artery, and cardiac chambers, while the standard variant was applied to remaining substructures. For Cohort II, nnU-Net and TotalSegmentator required manual reorientation to standard anatomical alignment for acceptable performance, whereas SMIT-based models processed scans in their native orientations. A model card summarizing all configurations is provided in Table 1.

**2. 6. Evaluation metrics and statistical analysis**

Segmentation accuracy was evaluated using the Hausdorff distance at the $95^{th}$ percentile (HD95) as the primary metric, with Dice similarity coefficient (DSC), surface DSC (sDSC, 1 mm tolerance), volume ratio (VR), and dose metrics as secondary metrics. Failed segmentations were assigned penalty values (HD95 = 50 mm, DSC = 0.0, sDSC = 0.1, and VR = 10) for fair comparison. Cases were excluded only when all models failed to detect a structure, thereby preventing selective reporting bias.

Statistical differences with respect to SMIT-Balanced were performed using two-sided Mann-Whitney U tests with Bonferroni correction for multiple comparisons. The influence of various factors on the model segmentation accuracy was evaluated using Spearman rank correlations (for age, and body mass index) and Mann-Whitney U tests (for gender, contrast and patient positioning). Segmentation metrics and statistical analyses were performed using SciPy (v1.11.1) and pandas (v2.2.2).

Dose-based evaluation was conducted by first calculating the dose distribution overlaid on the manual delineations within the dose map. The AI-segmentations were then overlaid on the same distribution, and dose-volume metrics, including maximum dose ($D_{Max}$) and mean dose ($D_{Mean}$), were computed for all structures. PA was evaluated separately using $V_{40}$ Gy [29]. This approach enabled a direct dose-based comparison between manual delineations and AI-segmentations.

## 3. Results

The evaluated cardiac substructures included the aorta (AO), pulmonary artery (PA), superior vena cava (SVC), inferior vena cava (IVC), right atrium (RA), right ventricle (RV), left atrium (LA), and left ventricle (LV).

### 3. 1. Comparison to adaptive architectures and open-source baselines

SMIT-Balanced was similarly accurate as nnU-Net (aggregate HD95: Cohort I: 5.5 $\pm$ 2.5 mm; Cohort II: 10.0 $\pm$ 9.4 mm). In Cohort I, nnU-Net showed lower HD95 for AO (3.0 $\pm$ 1.3 mm vs. 4.9 $\pm$ 5.5 mm, $p = 0.005$), and in Cohort II, nnU-Net showed lower HD95 for SVC ($p = 0.04$) (Figure 1).

No significant differences were observed for DSC, sDSC, or volume ratio for any substructure in Cohort I. In Cohort II, nnU-Net showed higher DSC for AO ($p = 0.02$), PA ($p=0.04$), SVC ($p =0.02$), and LV ($p = 0.02$), and improved volume ratio for AO ($p = 0.008$). Relative accuracy degradation from Cohort I to Cohort II was 62% for nnU-Net versus 50% for SMIT-Balanced.

TotalSegmentator showed higher errors across all substructures and metrics than SMIT-Balanced in Cohort I (14.8 $\pm$ 6.3 mm, $p < 0.001$) and Cohort II (31.6 $\pm$ 19.3 mm), corresponding to a 114% degradation. TotalSegmentator also failed to generate segmentations for SVC and IVC in nearly all scans, with only 1 out of 60 in Cohort I segmented for SVC (Figure 2).

### 3. 2. Comparison of balanced curriculum to oracle training

SMIT-Balanced was similarly accurate as SMIT-Oracle across all cardiac substructures despite using 64% fewer training scans. In Cohort I, the aggregate HD95 was 6.6 $\pm$ 4.3 mm for SMIT-Balanced versus 5.4 $\pm$ 2.6 mm for SMIT-Oracle, and in Cohort II, it was 10.0 $\pm$ 9.4 mm versus 9.4 $\pm$ 9.8 mm. Secondary metrics (DSC, sDSC, and volume ratio; Figures S3-S5) showed no differences except volume ratio in Cohort II ($p < 0.001$).

The relative accuracy degradation from Cohort I to Cohort II was lower for SMIT-Balanced (50%) than SMIT-Oracle (65%).

We further evaluated impact of different numbers of training sizes, which showed that the aggregate accuracy gap reduced after total of 64 scans (Figure S2). Performance improved notably from 16-16 to 32-32 scans (mean HD95 reduction of 0.66 mm across 5-fold cross-validation), whereas further increases beyond 32--32 yielded only marginal improvements (<0.3 mm per step) and were not statistically significant ($p > 0.05$), indicating performance saturation.

## 3. 3. Robustness to patient characteristics, imaging variations, and cross-domain generalization

### 3. 3. 1. Patient characteristics

SMIT-Balanced showed minimal correlation between segmentation accuracy (HD95) and patient age, with Spearman correlation coefficients (ρ) of -0.19 for AO to 0.12 for RA (Figure 3). For body mass index (BMI), a moderate negative correlation ($\rho = -0.33$) was observed only for PA, while other substructures remained weakly correlated (-0.11 to 0.09). No significant differences were observed across biological sex for any substructure (all $p > 0.05$).

SMIT-Oracle and nnU-Net demonstrated similarly weak correlations with age (ρ: -0.24 to 0.13) and BMI ($\rho = -0.27$ and -0.30 for PA, respectively; Figures S6-S7).

### 3. 3. 2. Imaging contrast

SMIT-Balanced showed differences between CECT and NCCT only for AO ($p = 0.021$), SVC ($p = 0.013$), and RA ($p = 0.024$), while the remaining five structures showed no differences (Figure S8). nnU-Net showed differences for AO ($p = 0.011$) and SVC ($p = 0.003$), and SMIT-Oracle differed only for IVC ($p = 0.024$).

Fine-tuning with frozen backbone resulted in similar accuracies for multiple sub-structures (Figure S9) except AO ($p = 0.015$), SVC ($p = 0.012$), and RA ($p = 0.006$). Training with NCCT-only reduced performance when evaluated on CECT across all substructures, while training with CECT-only reduced on NCCT-only for AO ($p = 0.005$).

### 3. 3. 3. Patient positioning

All models were affected by prone positioning, with larger effects observed for cardiac chambers than great vessels (Figure S10). SMIT-Balanced showed increased HD95 for IVC ($p = 0.003$), RA (p = 0.004), and RV (p = 0.007), LA (p = 0.016), and LV (p = 0.003). SMIT-Oracle showed increases for IVC (p = 0.014), RV (p < 0.001), and LV (p < 0.001), whereas nnU-Net differed for AO (p = 0.014), IVC (p < 0.001), and RV (p = 0.016).

### 3. 3. 4. Cross-data domain generalization

SMIT, trained on the TotalSegmentator dataset, showed lower HD95 for the AO, SVC, IVC, and LA (all p < 0.001), while nnU-Net-based TotalSegmentator showed lower error only for PA (p < 0.001). Performance for RA and LA was comparable between SMIT and nnU-Net-based TotalSegmentator.

On Cohort II, our model maintained significantly improved performance for SVC, IVC, and LV (all p < 0.0001), with no significant differences for AO, RV, or LA. TotalSegmentator showed lower HD95 only for PA (p < 0.001).

TotalSegmentator overall performance was 14.8 $\pm$ 6.3 mm (Cohort I) and 31.6 $\pm$ 19.3 mm (Cohort II). SVC and IVC segmentations were not generated in most scans across cohorts.

### 3. 4. Balanced model produced similar dose metrics as manual delineations

Mean dose correlation across substructures showed $r^2 > 0.95$. Quantitative DVH analysis demonstrated median absolute difference of 1.73 percentage points (IQR: 1.18-2.27) for Cohort I and 0.04 percentage points (IQR: 0.01-0.20) for Cohort II (Figure 4), indicative of excellent agreement between AI-segmentations and manual delineations (Figure 5, S11). The mean dose differences were below 1 Gy for all cardiac chambers, and volume metrics showed no statistically significant differences (all p > 0.05, Tables S1 and S2).

SMIT-Oracle and nnU-Net achieved similar DVH performance (Figure S12).

## 4. Discussion

This study developed and validated a hybrid pretrained transformer-convolutional network called SMIT to segment cardiac substructures from CT scans across common imaging variations. SMIT's accuracy was comparable to nnU-Net while requiring lesser training data iterations. It also required substantially fewer labeled training data than the oracle model to achieve similar performance. Performance was robust to imaging variations and patient positioning. Our approach allowed for auto-segmentation of volumetric CT scans without task-specific architectural modifications.

Our results showed that pretraining without data-specific network reconfiguration was sufficient to achieve as good accuracy as task-specific reconfigured nnU-Net. Although aggregate segmentation accuracy was comparable between SMIT and nnU-Net, the pretrained model demonstrated smaller performance degradation across cohorts acquired under different clinical conditions, suggesting reduced dependence on training-data-specific characteristics. SMIT also showed slightly improved performance on cross-domain data than nnU-Net-based TotalSegmentator [23,28]. The latter approach as well as in-house trained nnU-Net required manual scan reorientation to canonical anatomical alignment for stable inference on the evaluated held-out cohort, precluding possibility to produce batched and automated analysis with these methods.

Multiple studies have explored automated segmentation of cardiac substructures using atlas-based and deep learning approaches. Unlike atlas-based methods (e.g., Finnegan et al. [13] and Oevert et al. [14]), which rely on multi-step pipelines requiring prior lung and/or heart segmentation, our approach avoided serialized preprocessing steps that can propagate upstream errors and complicate clinical commissioning and quality assurance. This shift aligned with broader adoption of deep learning, which demonstrated superior performance compared with commercial atlas-based techniques [30]. Previous deep learning work using 2D DeepLab [9] focused only on supine lung cancer cohorts and did not evaluate variability across contrast-enhanced and non-contrast CT or other patient factors. In contrast, large-scale clinical validations showed that modern deep learning models substantially reduced segmentation time while maintaining or exceeding expert accuracy and remained robust to common clinical challenges such as imaging artifacts and non-contrast scans typical in breast radiotherapy workflows, thereby improving both efficiency and consistency of volumetric cardiac segmentation across diverse patient populations [31-32].

The strengths of this study include evaluation of pretraining-based models across multiple clinically relevant imaging variations and assessment of model transferability across patient populations without retraining. Ablation analysis demonstrated that frozen-encoder training performed similarly to full fine-tuning, indicating that task-relevant anatomical representations were largely captured during pretraining. Our study had a few limitations. The evaluation was restricted to a single institution due to patient privacy concerns, and multi-institutional validation is planned next step to establish generalizability. Second, the current study did not undertake an evaluation of potential reduction in contouring time by clinicians, rather focusing on robustness across imaging and patient variations. Coronary arteries and cardiac valves were not included, and the use of single-expert delineations precluded assessment of inter-observer variability. Nevertheless, our work provides unique insights into the benefits of using pretrained transformer-based network formulations for generating automated cardiac substructures segmentation that included data requirements, robustness analysis, and a deployment friendly automated workflow.

Table 1: Model card summarizing all evaluated segmentation models.

| Model name | Training data | Input crop size | Training iterations | Purpose |
|---|---|---|---|---|
| *Transformer pretrained on CT scans using self-supervised learning (SMIT) with U-Net decoder* | | | | |
| SMIT-Balanced | Cohort I [32 CECT 32 NCCT] | $128^3$ | 32,000 | Data efficiency |
| SMIT-Oracle | Cohort I [56 CECT 124 NCCT] | $128^3$ | 32,000 | Upper bound |
| *Ablations for SMIT models* | | | | |
| Frozen backbone | Cohort I [32 CECT 32 NCCT] | $128^3$ | 32,000 | Pretrained backbone ablation |
| CECT-only | Cohort I [56 CECT] | $128^3$ | 32,000 | Contrast ablation |
| NCCT-only | Cohort I [56 NCCT] | $128^3$ | 32,000 | Contrast ablation |
| SMIT-TotalSegmentator | Dataset accompanied with TotalSegmentator | $128^3$ | 32,000 | Data generalization ablation |
| *Self-configuring nnU-Net* | | | | |
| nnU-Net | Cohort I [32 CECT 32 NCCT] | 192 x 192 x 64 | 250,000 | Self-configuring CNN comparison |
| TotalSegmentator | Dataset accompanied with TotalSegmentator | [Open-sourced weights and code] | | Open-sourced CNN comparison for superior vena cava and inferior vena cava |
| TotalSegmentator *heartchambers_highres* | Updated dataset accompanied with TotalSegmentator | [Open-sourced weights and code] | | Open-sourced CNN comparison for aorta, pulmonary artery, and cardiac chambers |

## Figures

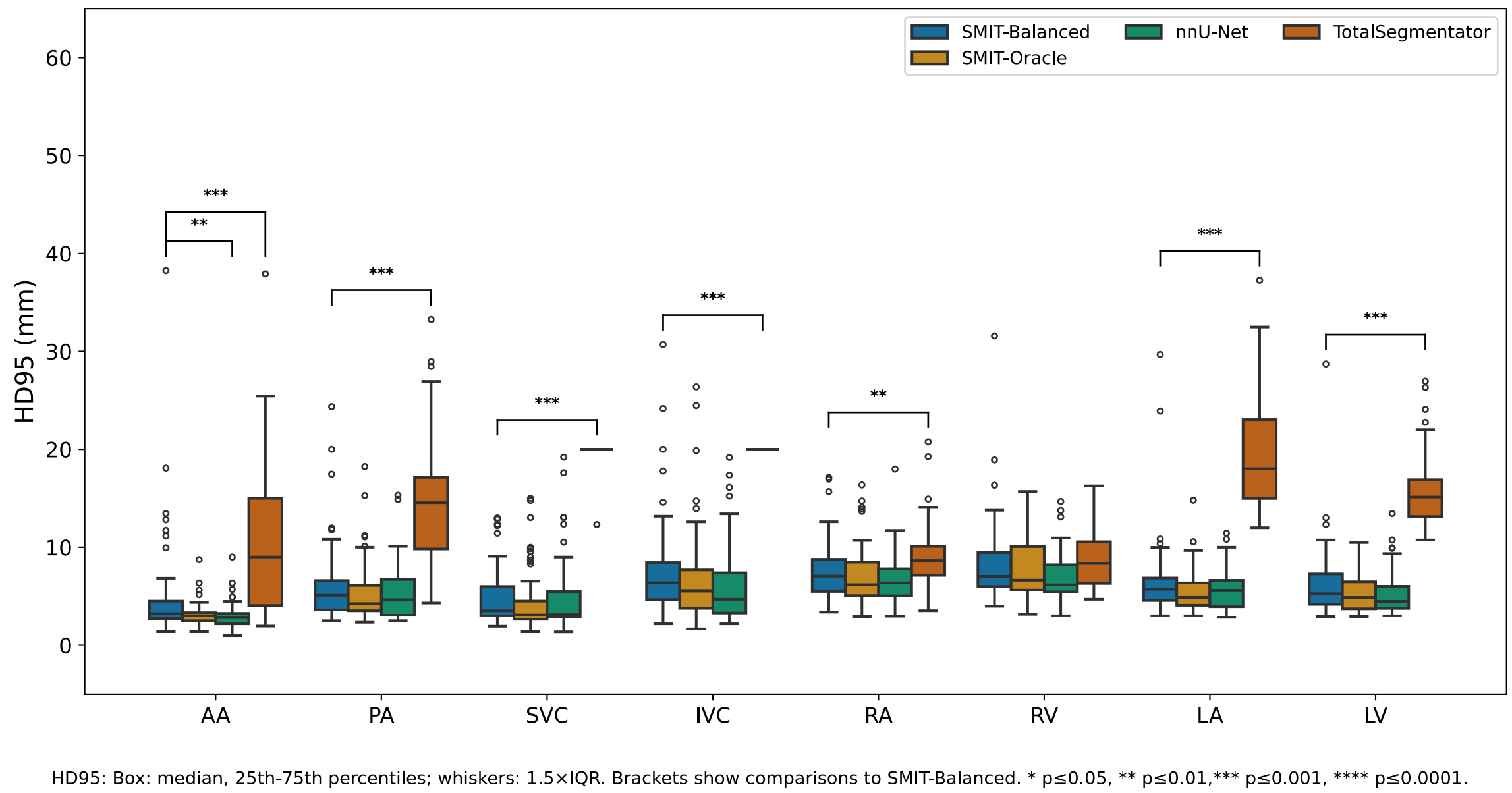

(a)

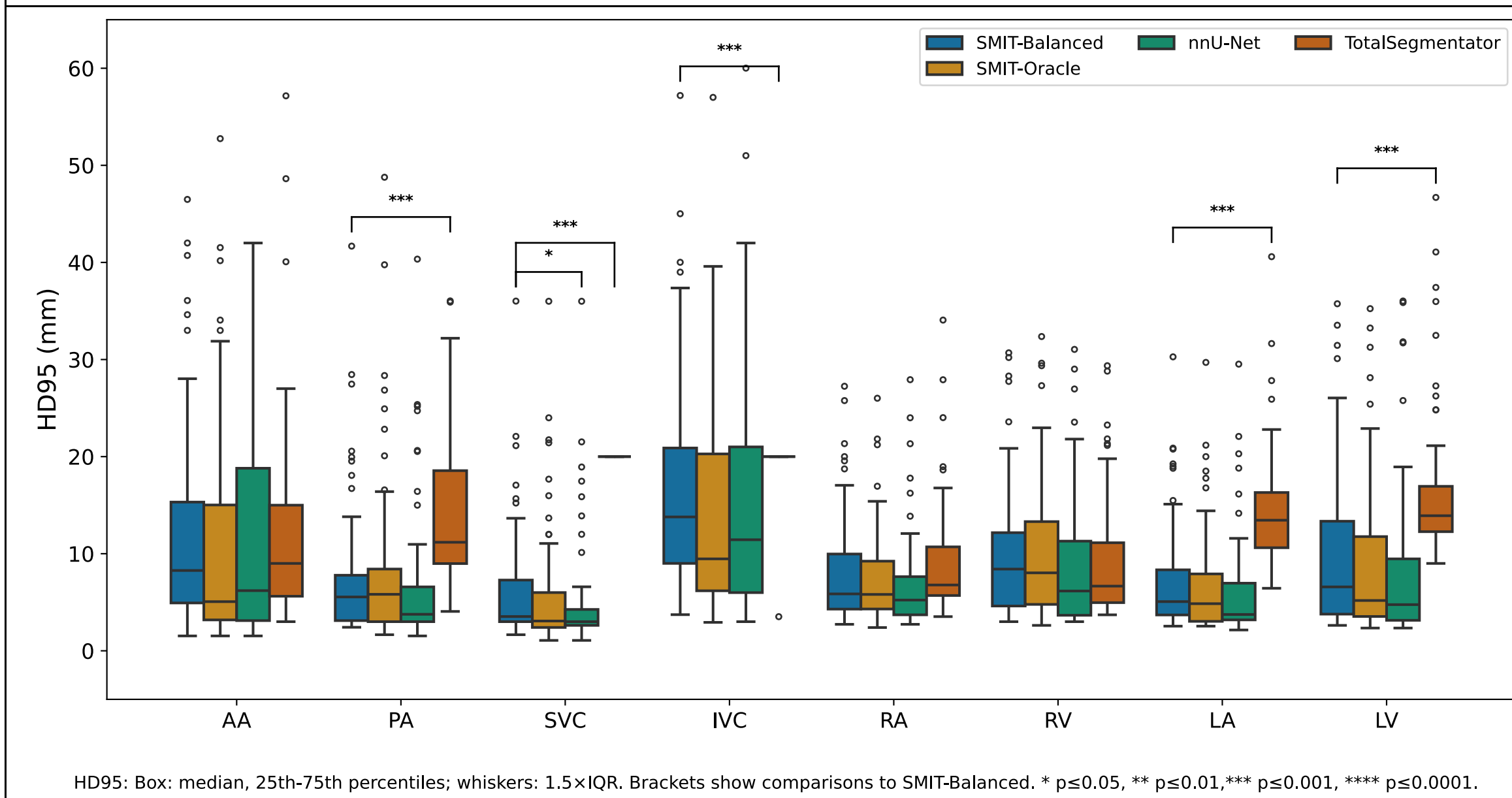

(b)

Figure 1: Segmentation performance quantified with Hausdorff Distance at 95th percentile (HD95) for (a) Cohort I and (b) Cohort II.

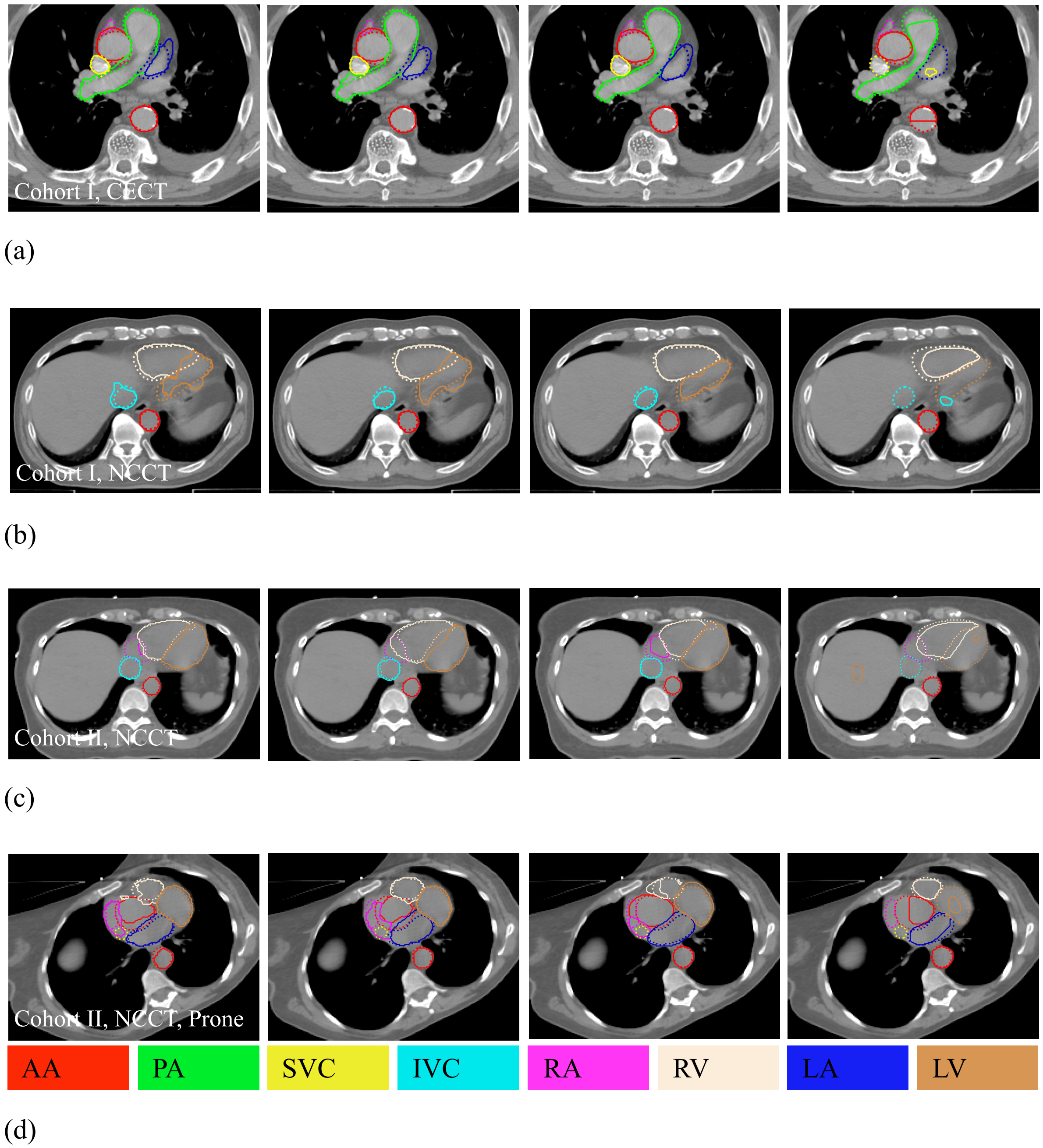

Figure 2: Representative auto-segmentation results across imaging conditions. Scans (a–d) from Cohorts I and II showing CECT and NCCT acquisitions. Segmentations from left to right – SMIT-Balanced, SMIT-Oracle, nnU-Net, and TotalSegmentator.

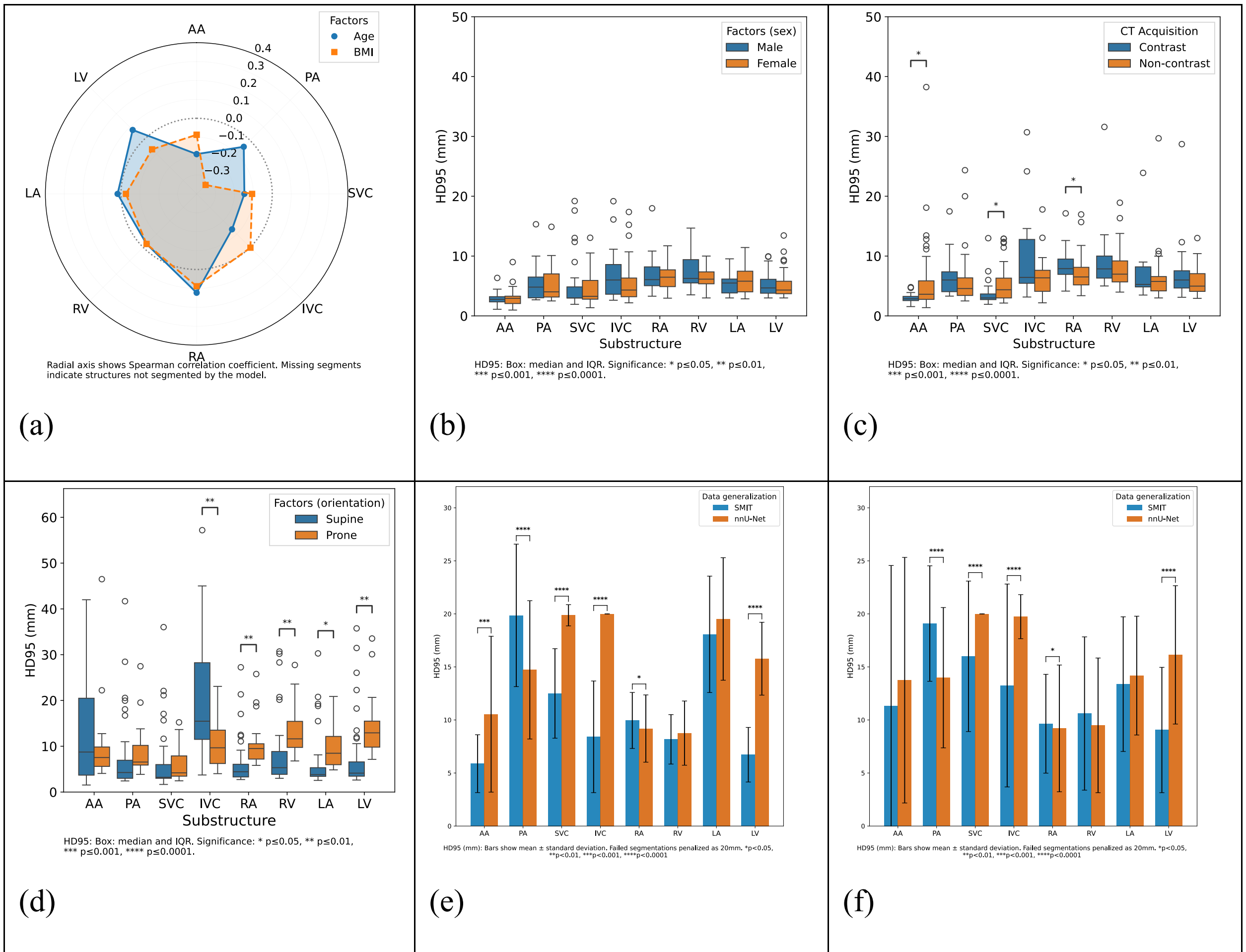

Figure 3: Robustness analysis. (a-d) Segmentation performance of the Balanced model across variations in (a) age and body mass index (BMI), (b) biological sex, (c) presence of intravenous contrast, and (d) scanning position. Analyses (a–c) used Cohort I; analysis (d) used Cohort II. (e-f) Domain generalization of SMIT and nnU-Net when trained on TotalSegmentator dataset and evaluated on cohort I and cohort II.

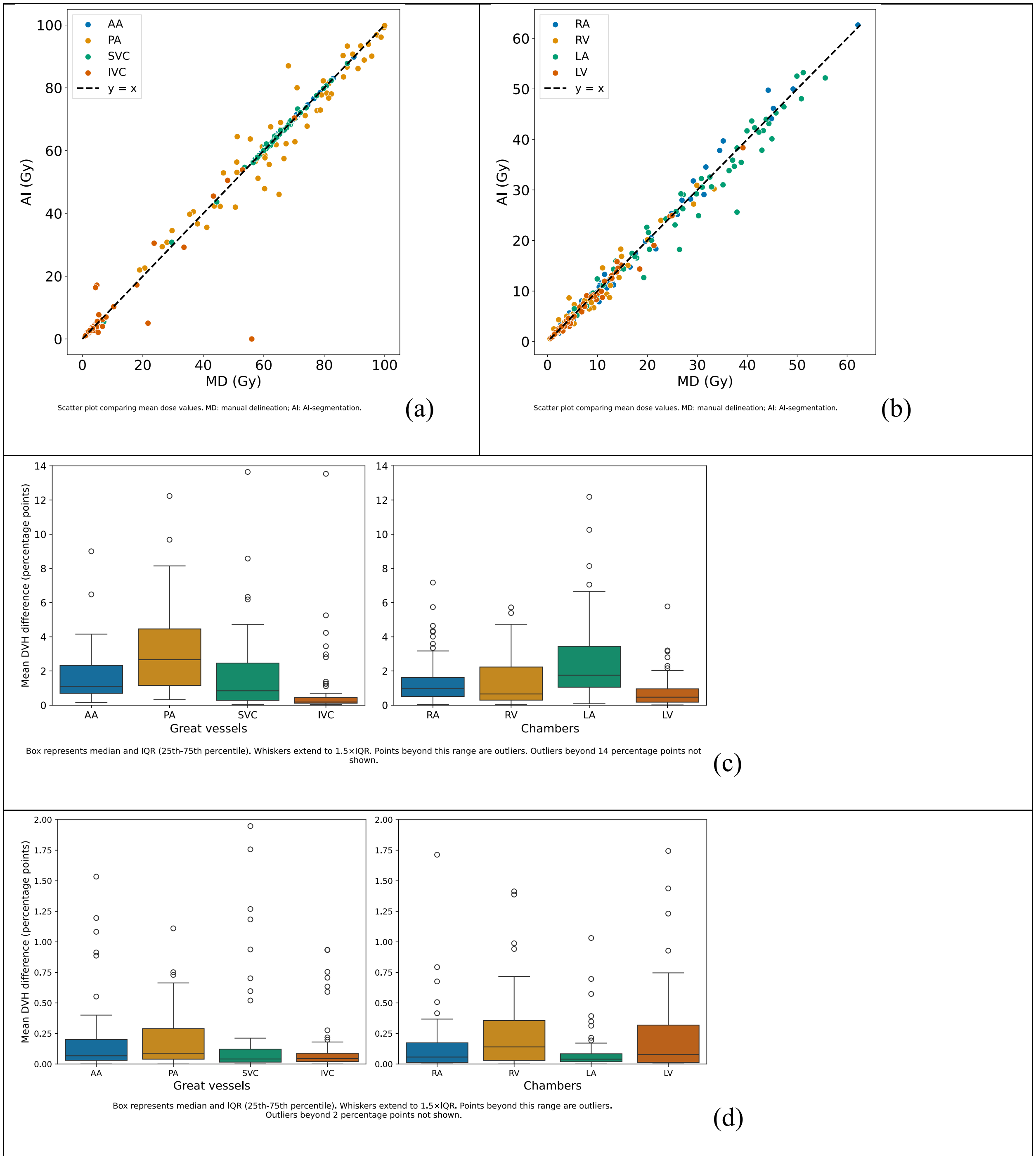

Figure 4: Dose-volume metric validation comparing AI-segmentation to manual delineations. (a, b) Scatter plots between manual delineations (MD) and auto-segmentations (AI) from SMIT-Balanced for different cardiac substructures in cohort I. The dashed line represents the identity line (y = x), indicating perfect agreement. (c) DVH difference (Mean absolute error, %) for cohort I and (d) for cohort II.

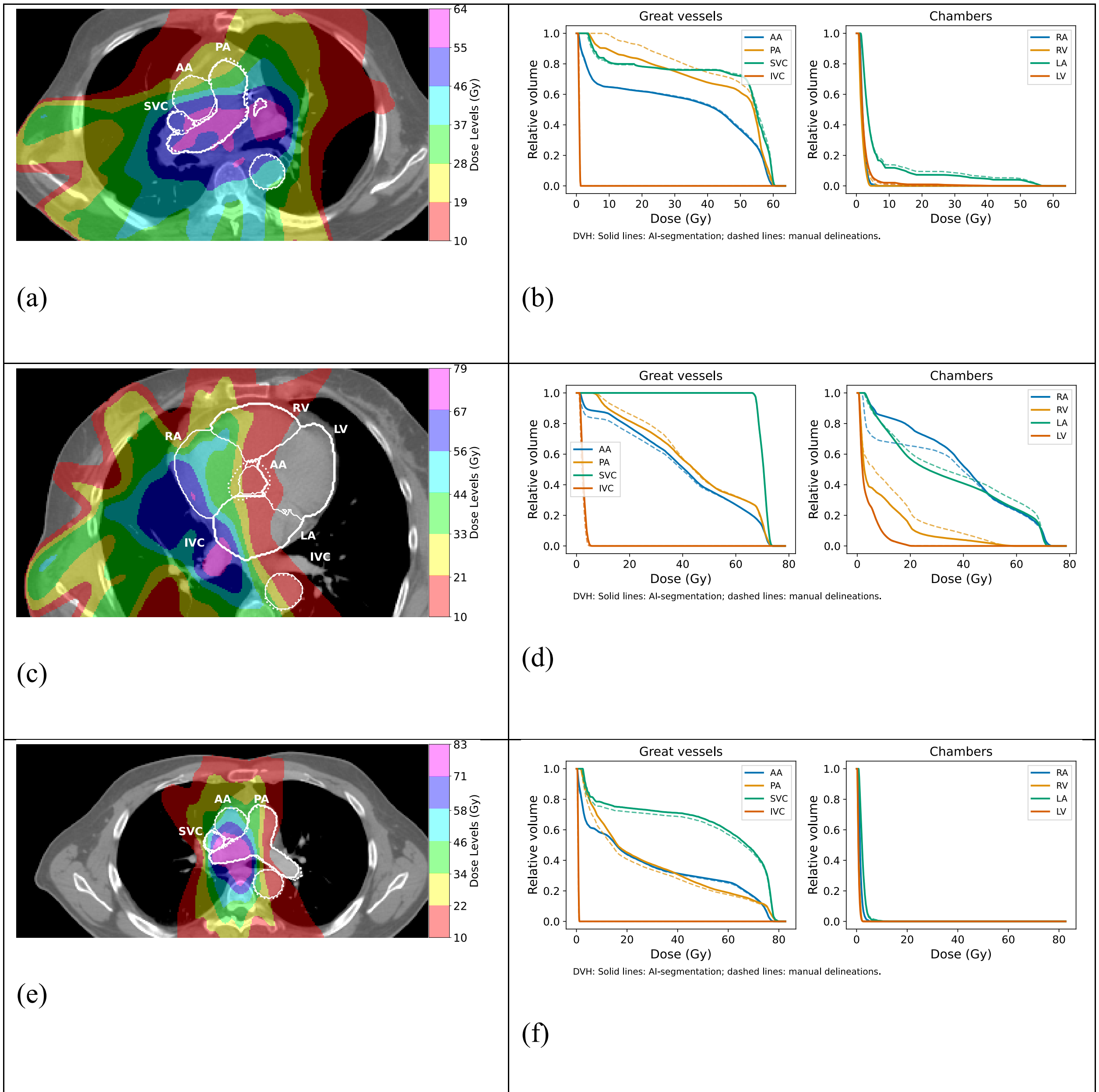

Figure 5: Dose-volume metric comparison. Radiation dose distributions (a, c, e) and corresponding dose–volume histograms (DVHs) (b, d, f) for SMIT-Balanced model on three representative scans from Cohort I. Solid lines indicate model segmentations; dotted lines indicate manual delineations.

**Acknowledgements**

This research was partially supported by the NCI R01CA258821 and the Memorial Sloan Kettering Cancer Center Support Grant/Core Grant NCI P30CA008748.

**References**

[1] Friedes C, Iocolano M, Lee SH, Duan L, Li B, Doucette A, et al. The effective radiation dose to immune cells predicts lymphopenia and inferior cancer control in locally advanced NSCLC. Radiother Oncol. 2024;190:110030. https://doi.org/10.1016/j.radonc.2023.110030.

[2] Xu C, Jin JY, Zhang M, Liu A, Kong FS, Lin SH, et al. The impact of the effective dose to immune cells on lymphopenia and survival of esophageal cancer after chemoradiotherapy. Radiother Oncol. 2020;146:180-6. https://doi.org/10.1016/j.radonc.2020.02.015.

[3] Yu Y, Fu P, Jin JY, Gao S, Wang W, Machtay M, et al. Impact of effective dose to immune cells (EDIC) on lymphocyte nadir and survival in limited-stage SCLC. Radiother Oncol. 2021;162:26-33. https://doi.org/10.1016/j.radonc.2021.06.020.

[4] Stoltzfus KC, Zhang Y, Sturgeon K, Sinoway LI, Trifiletti DM, Chinchilli VM, et al. Fatal heart disease among cancer patients. Nat Commun. 2020;11:2011. https://doi.org/10.1038/s41467-020-15639-5.

[5] Bowen Jones S, Marchant T, Saunderson C, McWilliam A, Banfill K. Moving beyond mean heart dose: The importance of cardiac substructures in radiation therapy toxicity. J Med Imaging Radiat Oncol. 2024;68(8):974-86. https://doi.org/10.1111/1754-9485.13737.

[6] Isensee F, Jaeger PF, Kohl SA, Petersen J, Maier-Hein KH. nnU-Net: a self-configuring method for deep learning-based biomedical image segmentation. Nat Methods. 2021;18(2):203-11. https://doi.org/10.1038/s41592-020-01008-z.

[7] Saha M, Jung JW, Lee SW, Lee C, Lee C, Mille MM. A deep learning segmentation method to assess dose to organs at risk during breast radiotherapy. Phys Imaging Radiat Oncol. 2023. https://doi.org/10.1016/j.phro.2023.100520.

[8] Carles M, Kuhn D, Fechter T, Baltas D, Mix M, Nestle U, et al. Development and evaluation of two open-source nnU-Net models for automatic segmentation of lung tumors on PET and CT

images with and without respiratory motion compensation. Eur Radiol. 2024; 34:6701-11. https://doi.org/10.1007/s00330-024-10751-2.

[9] Haq R, Hotca A, Apte A, Rimner A, Deasy JO, Thor M. Cardio-pulmonary substructure segmentation of radiotherapy computed tomography images using convolutional neural networks for clinical outcomes analysis. Phys Imaging Radiat Oncol. 2020 Jun 10;14:61-66. https://doi.org/10.1016/j.phro.2020.05.009.

[10] Morris ED, Ghanem AI, Dong M, Pantelic MV, Walker EM, Glide-Hurst CK. Cardiac substructure segmentation with deep learning for improved cardiac sparing. Med Phys. 2020. https://doi.org/10.1002/mp.13940.

[11] Harms J, Lei Y, Tian S, McCall NS, Higgins KA, Bradley JD, et al. Automatic delineation of cardiac substructures using a region-based fully convolutional network. Med Phys. 2021. https://doi.org/10.1002/mp.14810.

[12] Chen X, Mumme RP, Corrigan KL, Mukai-Sasaki Y, Koutroumpakis E, Palaskas NL, et al. Deep learning–based automatic segmentation of cardiac substructures for lung cancers. Radiother Oncol. 2024. https://doi.org/10.1016/j.radonc.2023.110061.

[13] Finnegan RN, Chin V, Chlap P, Haider A, Otton J, Dowling J, et al. Open-source, fully automated hybrid cardiac substructure segmentation: Development and optimisation. Phys Eng Sci Med. 2023;46(1):377–93. https://doi.org/10.1007/s13246-023-01231-w.

[14] Van Den Oever LB, Spoor DS, Crijns APG, Vliegenthart R, Oudkerk M, Veldhuis RNJ, et al. Automatic cardiac structure contouring for small datasets with cascaded deep learning models. J Med Syst. 2022;46(5):22. https://doi.org/10.1007/s10916-022-01810-6.

[15] Jiang J, Rangnekar A, Veeraraghavan H. Self-supervised learning improves robustness of deep learning lung tumor segmentation models to CT imaging differences. Med Phys. 2024. https://doi.org/10.1002/mp.17541.

[16] Jiang J, Tyagi N, Tringale K, Crane C, Veeraraghavan H. Self-supervised 3D anatomy segmentation using self-distilled masked image transformer (SMIT). Medical Image Computing and Computer-Assisted Intervention – MICCAI 2022. Cham: Springer Nature Switzerland; 2022:556–66. https://doi.org/10.1007/978-3-031-16440-8_53.

[17] Liu Z, Lin Y, Cao Y, Hu H, Wei Y, Zhang Z, et al. Swin transformer: Hierarchical vision transformer using shifted windows. Proceedings of the IEEE/CVF International Conference on Computer Vision. 2021:10012–22. https://doi.org/10.48550/arXiv.2103.14030.

[18] Thor M, Deasy J, Iyer A, Bendau E, Fontanella A, Apte A, et al. Toward personalized dose-prescription in locally advanced non-small cell lung cancer: validation of published normal tissue complication probability models. Radiother Oncol. 2019; 138:45-51. https://doi.org/10.1016/j.radonc.2019.05.011

[19] Thor M, Montovano M, Hotca A, Luo L, Jackson A, Wu AJ, et al. Are unsatisfactory outcomes after concurrent chemoradiotherapy for locally advanced non-small cell lung cancer due to treatment-related immunosuppression? Radiother Oncol. 2020;143:51-7. https://doi.org/10.1016/j.radonc.2019.07.016

[20] Hotca A, Thor M, Deasy JO, Rimner A. Dose to the cardio-pulmonary system and treatment-induced electrocardiogram abnormalities in locally advanced non-small cell lung cancer. Clin Transl Radiat Oncol. 2019;19:96-102. https://doi.org/10.1016/j.ctro.2019.09.003

[21] Mohan R, Barest G, Brewster LJ, Chui CS, Kutcher GJ, Laughlin JS, Fuks Z. A comprehensive three-dimensional radiation treatment planning system. Int J Radiat Oncol Biol Phys. 1988;15(2):481-95. https://doi.org/10.1016/s0360-3016(98)90033-5

[22] Thor M, Scott JM, Reiner AS, Hong L, Yu AF, Apte A, et al. Cardiopulmonary substructure doses are not correlated with cardiorespiratory fitness among breast cancer survivors treated with contemporary radiation therapy. Adv Radiat Oncol. 2024;9(12):101613. https://doi.org/10.1016/j.adro.2024.101613.

[23] Wasserthal J, Breit HC, Meyer MT, Pradella M, Hinck D, Sauter AW, et al. TotalSegmentator: robust segmentation of 104 anatomic structures in CT images. Radiology Artif Intell. 2023. https://doi.org/10.1148/ryai.230024.

[24] Feng M, Moran JM, Koelling T, Chughtai A, Chan JL, Freedman L, et al. Development and validation of a heart atlas to study cardiac exposure to radiation following treatment for breast cancer. Int J Radiat Oncol Biol Phys. 2011;79(1):10-8. https://doi.org/10.1016/j.ijrobp.2009.10.058.

[25] Ronneberger O, Fischer P, Brox T. U-net: Convolutional networks for biomedical image segmentation. Medical Image Computing and Computer-Assisted Intervention – MICCAI 2015. Vol 9351. Springer; 2015:234–41. https://doi.org/10.1007/978-3-319-24574-4_28.

[26] Cardoso MJ, Li W, Brown R, Ma N, Kerfoot E, Wang Y, et al. MONAI: An open-source framework for deep learning in healthcare. arXiv. 2022. https://doi.org/10.48550/arXiv.2211.02701.

[27] Paszke A, Gross S, Massa F, Lerer A, Bradbury J, Chanan G, et al. PyTorch: An imperative style, high-performance deep learning library. Advances in Neural Information Processing Systems. 2019. https://doi.org/10.48550/arXiv.1912.01703.

[28] Hinck D, Segeroth M, Miazza J, Berdajs D, Bremerich J, Wasserthal J, et al. Automatic segmentations of cardiovascular structures on chest CT datasets: An update of the TotalSegmentator. Eur J Radiol. 2025. https://doi.org/10.1016/j.ejrad.2025.112006.

[29] Harris WB, Zou W, Cheng C, Jain V, Teo BKK, Dong L, et al. Higher dose volumes may be better for evaluating radiation pneumonitis in lung proton therapy patients compared with traditional photon-based dose constraints. Adv Radiat Oncol. 2020. https://doi.org/10.1016/j.adro.2020.06.023.

[30] Choi MS, Choi BS, Chung SY, Kim N, Chun J, Kim YB, et al. Clinical evaluation of atlas- and deep learning-based automatic segmentation of multiple organs and clinical target volumes for breast cancer. Radiother Oncol. 2020 Dec 1;153:139-45. https://doi.org/10.1016/j.radonc.2020.09.045.

[31] Zeleznik R, Weiss J, Taron J, Guthier C, Bitterman DS, Hancox C, et al. Deep-learning system to improve the quality and efficiency of volumetric heart segmentation for breast cancer. NPJ Digit Med. 2021 Mar 5;4(1):43. https://doi.org/10.1038/s41746-021-00416-5.

[32] Jin X, Thomas MA, Dise J, Kavanaugh J, Hilliard J, Zoberi I, et al. Robustness of deep learning segmentation of cardiac substructures in noncontrast computed tomography for breast cancer radiotherapy. Med Phys. 2021 Nov;48(11):7172-88. https://doi.org/10.1002/mp.15237.

## Supplemental

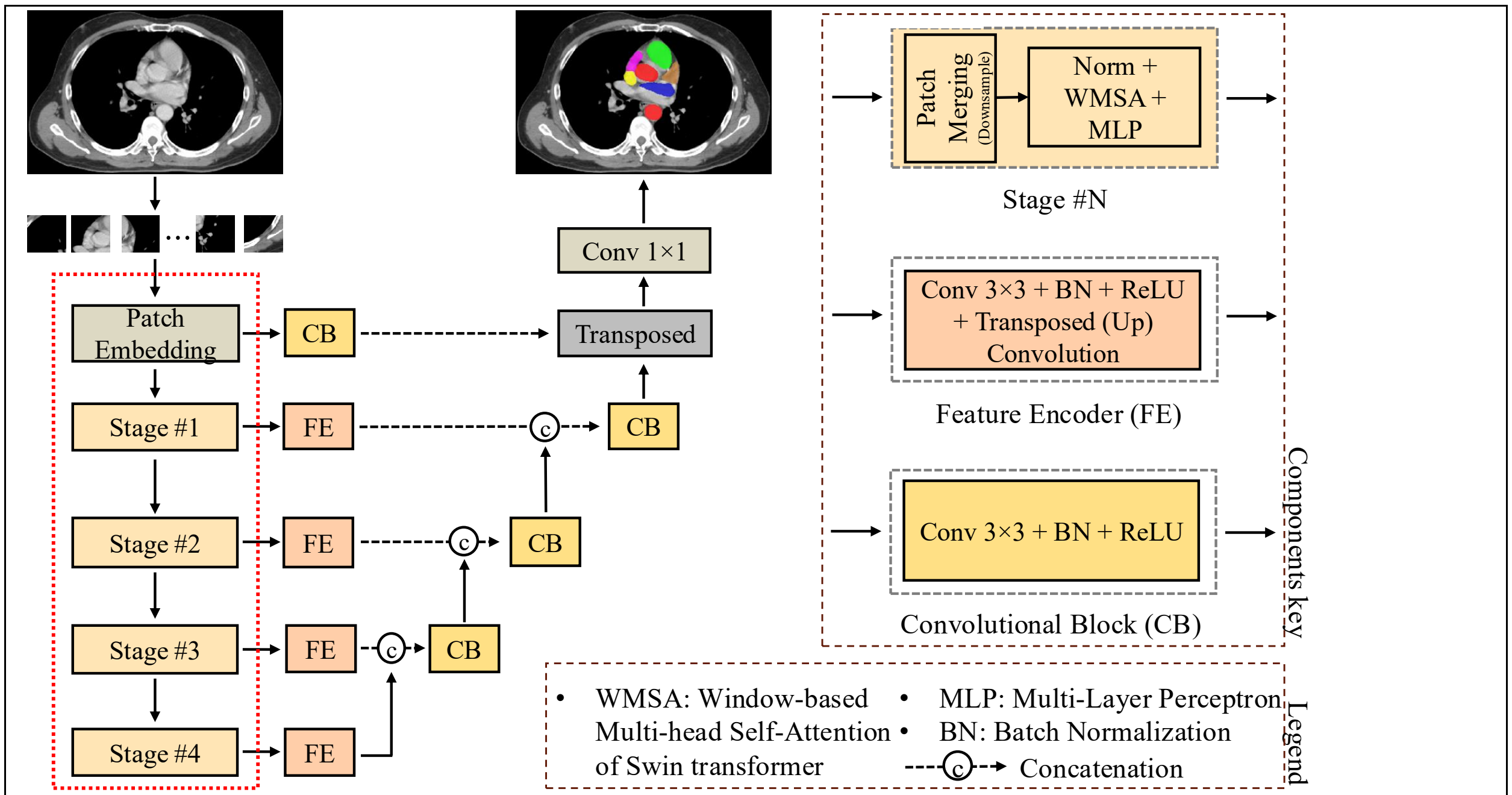

Supplementary Figure S1: Architectural diagram. The proposed segmentation network combines a pretrained Swin Transformer backbone encoder (indicated by the dotted red box) with the U-Net convolutional decoder.

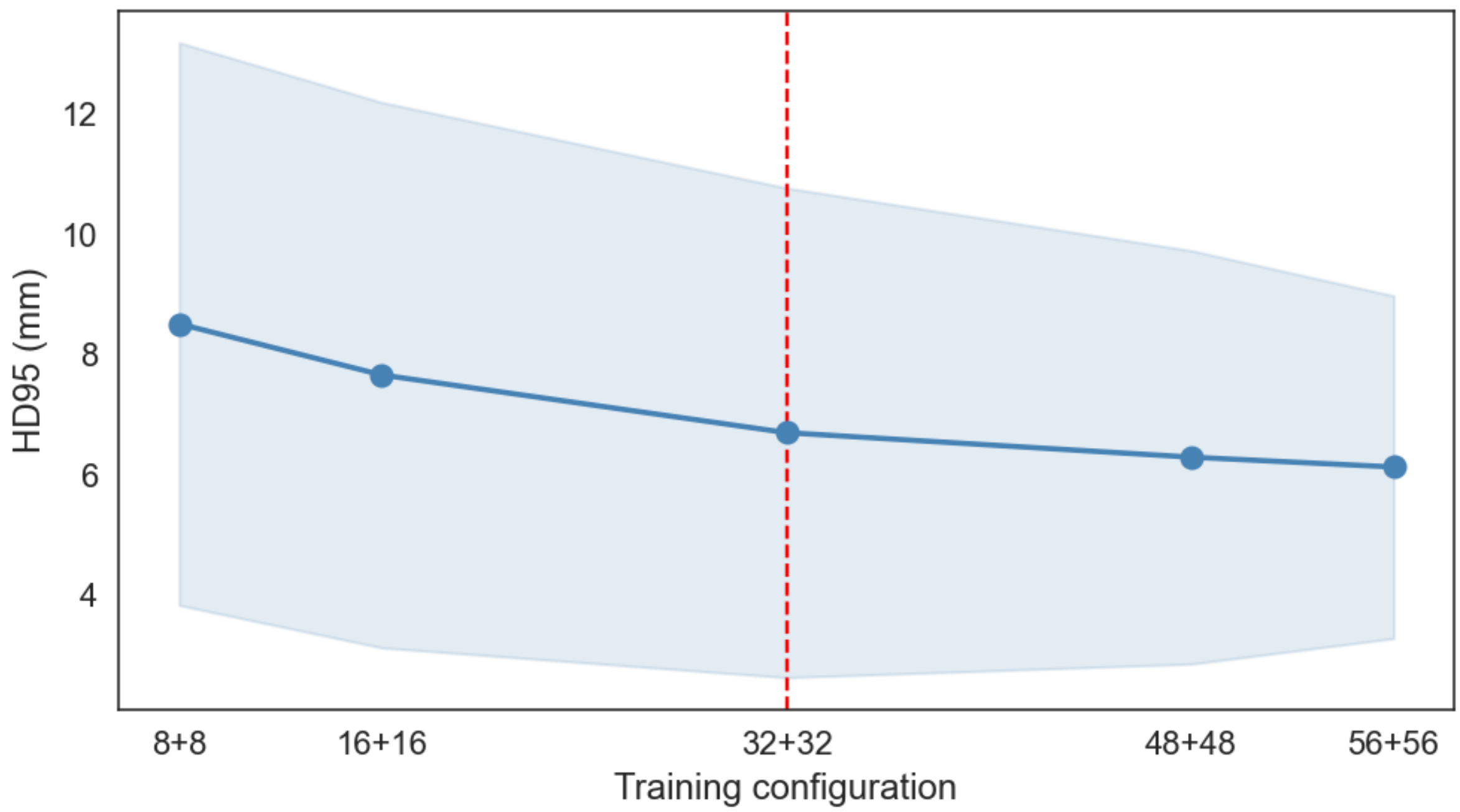

Shaded region indicates aggregate mean ± standard deviation. The dashed red line marks the 32+32 configuration, beyond which marginal improvement diminishes.

Supplementary Figure S2: Effect of training set size on accuracy of segmenting cardiac substructures. Configurations denote lung cancer (cohort I) training samples and their corresponding five-fold cross-validation scores.

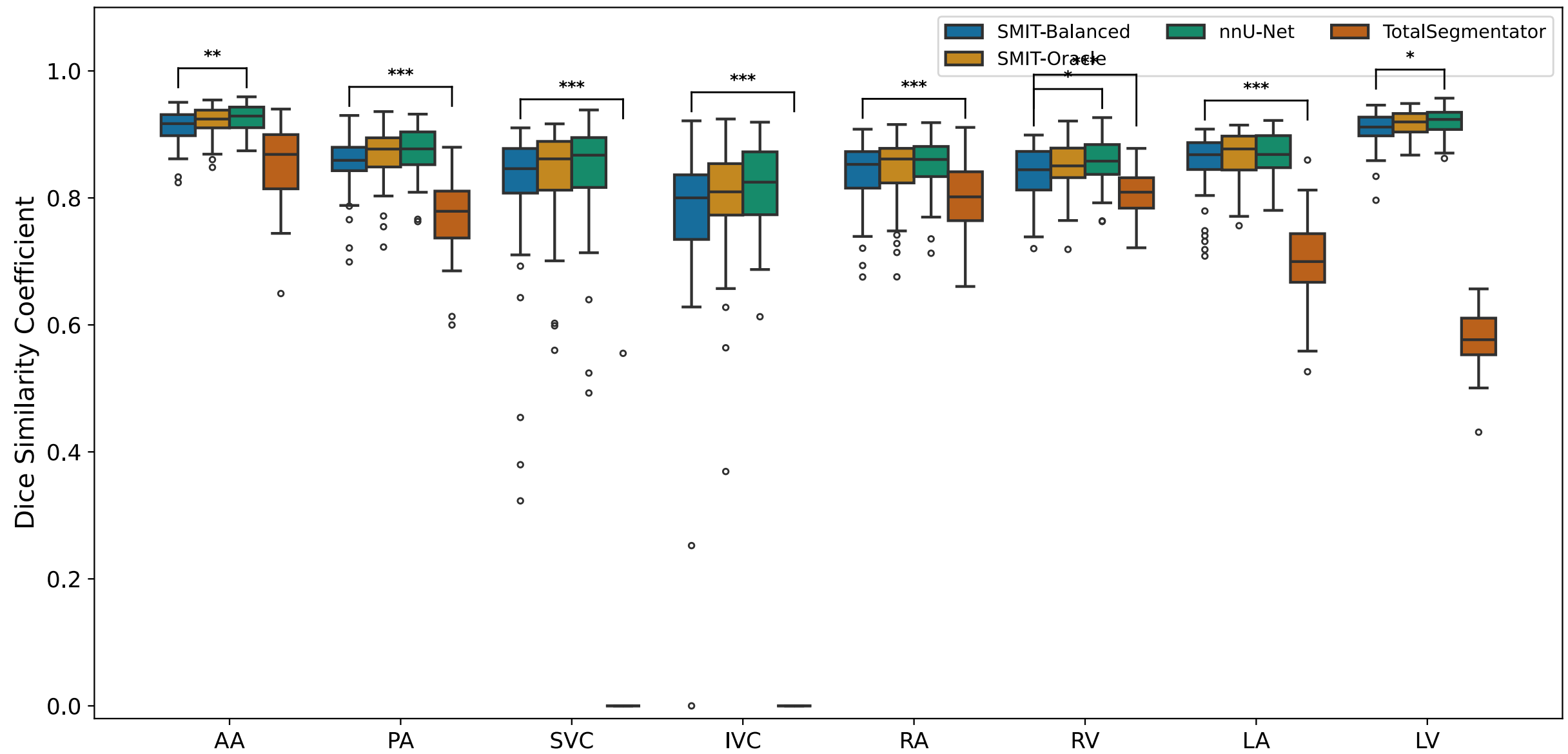

DSC: Box: median, 25th-75th percentiles; whiskers: 1.5×IQR. Brackets show comparisons to SMIT-Balanced.* p≤0.05, ** p≤0.01,*** p≤0.001, **** p≤0.0001.

(a)

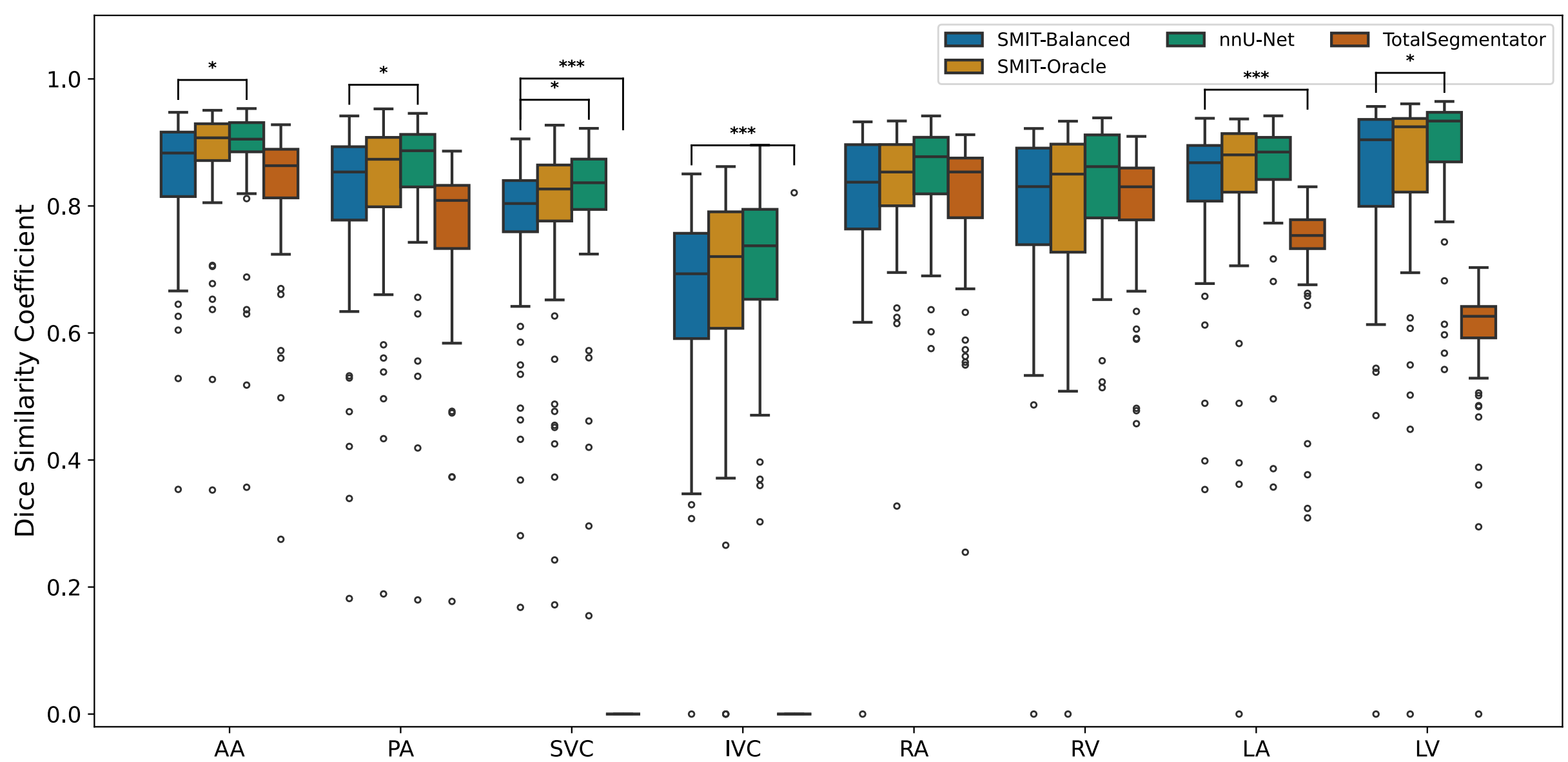

DSC: Box: median, 25th-75th percentiles; whiskers: 1.5×IQR. Brackets show comparisons to SMIT-Balanced.* p≤0.05, ** p≤0.01,*** p≤0.001, **** p≤0.0001.

(b)

Supplementary Figure S3: Segmentation performance quantified with Dice similarity coefficient (DSC) for Cohort I (panel a) and Cohort II (b).

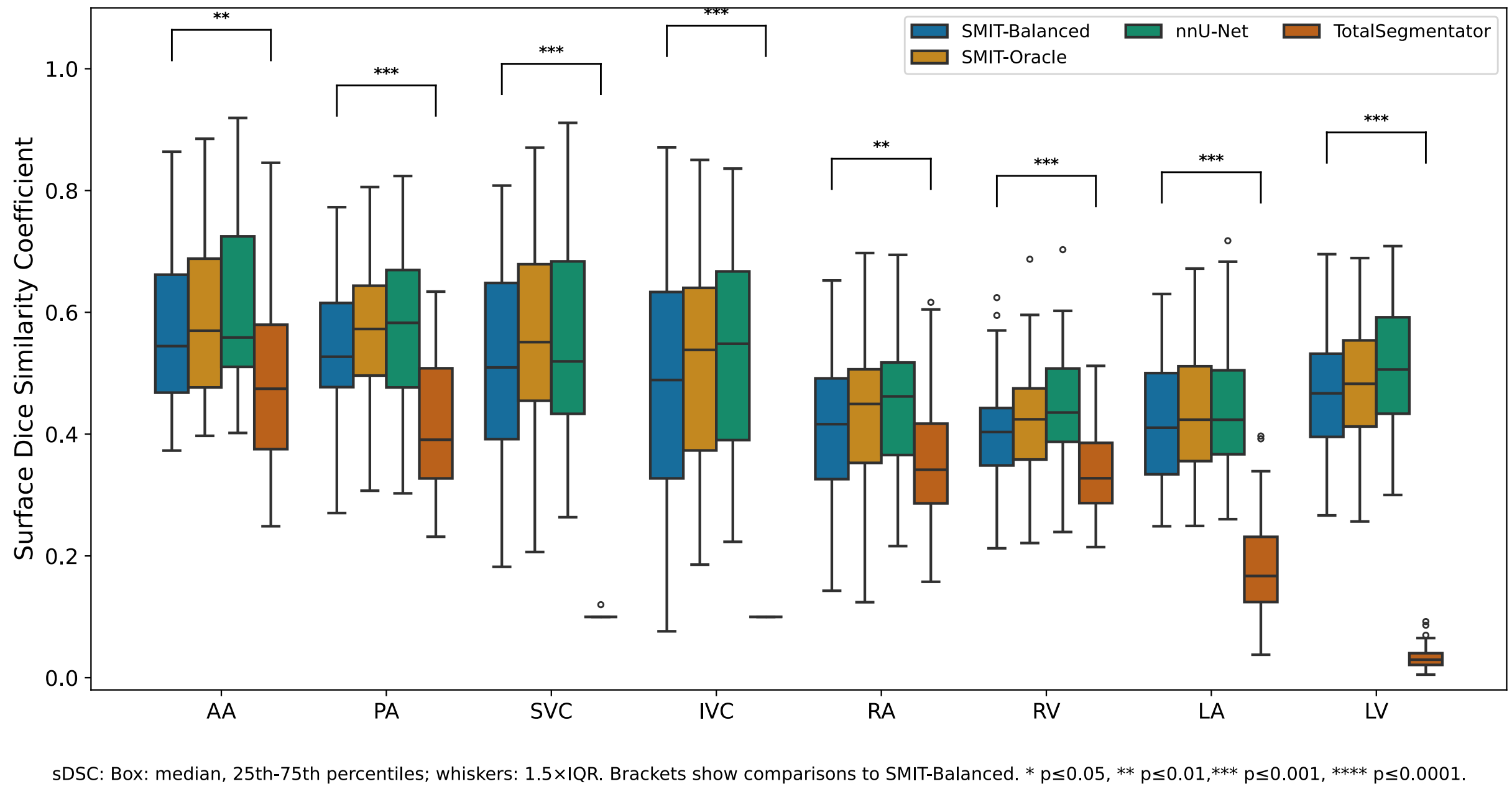

(a)

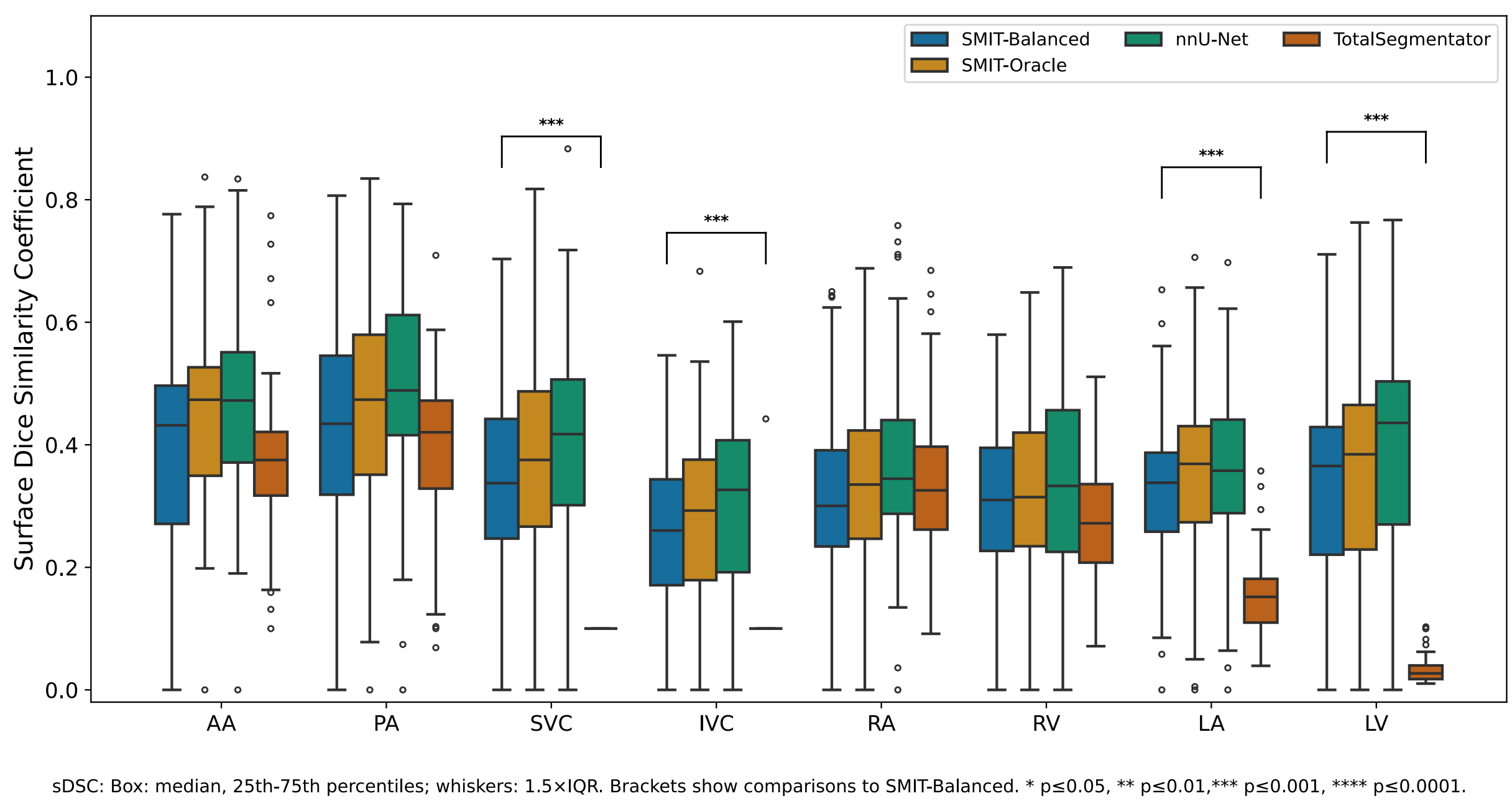

(b)

Supplementary Figure S4: Segmentation performance quantified with surface DSC (sDSC) for Cohort I (panel a) and Cohort II (b).

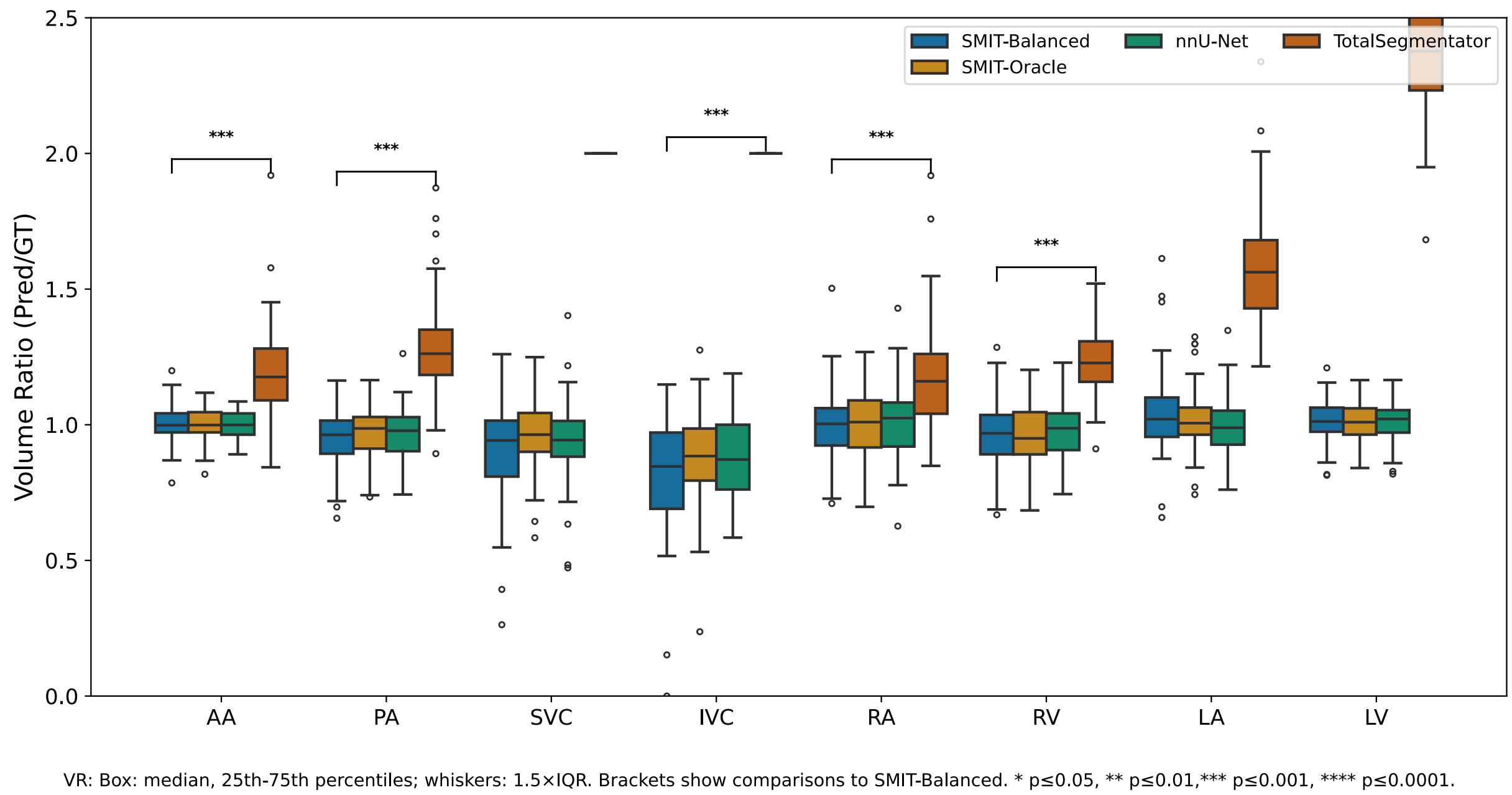

(a)

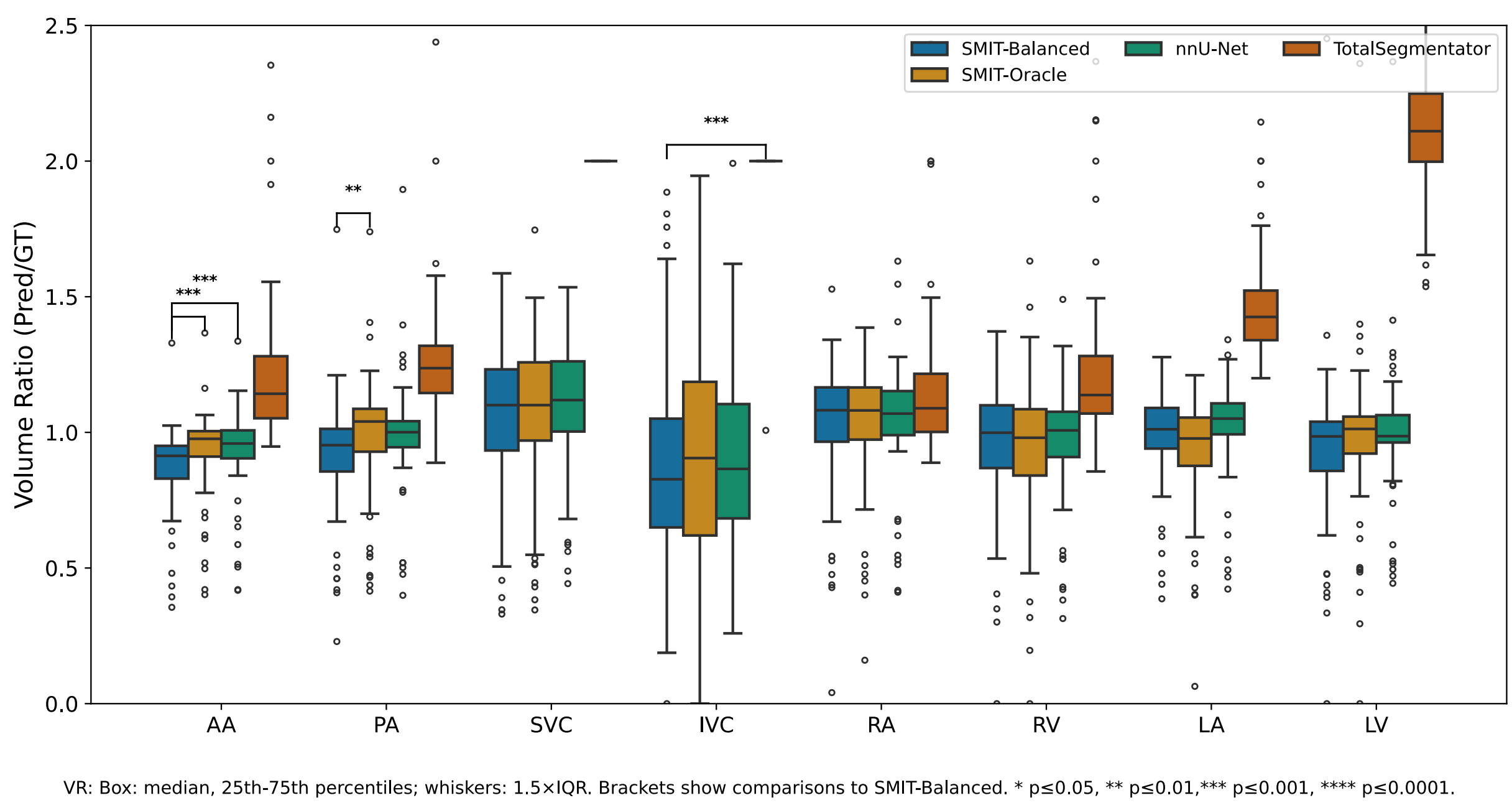

(b)

Supplementary Figure S5: Segmentation performance quantified with volume ratio for Cohort I (panel a) and Cohort II (b).

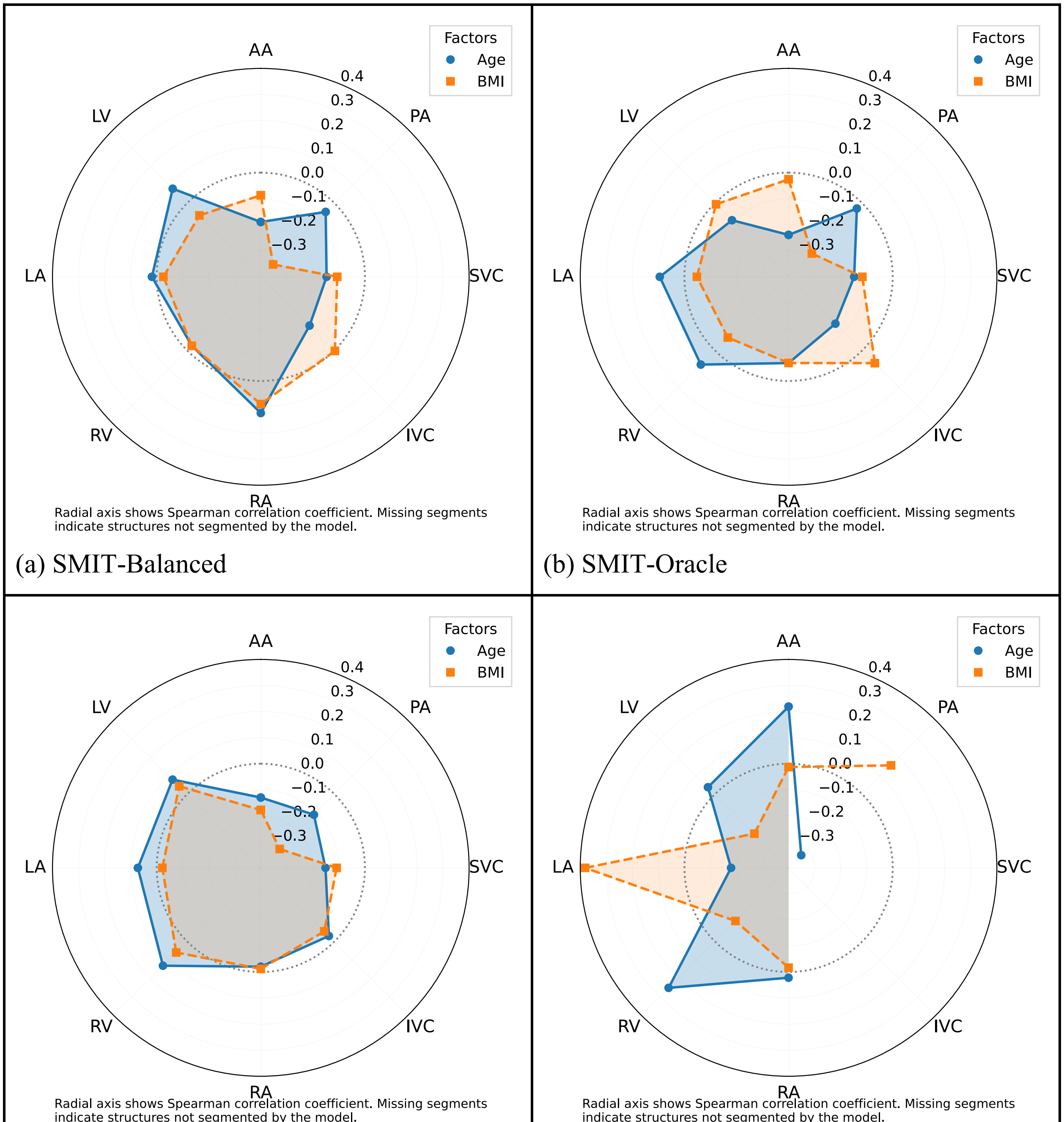

(a) SMIT-Balanced

(b) SMIT-Oracle

(c) nnU-Net-Balanced

(d) TotalSegmentator

Supplementary Figure S6: Robustness to age and body mass index for scans from cohort I. Radial axis shows Spearman correlation coefficient.

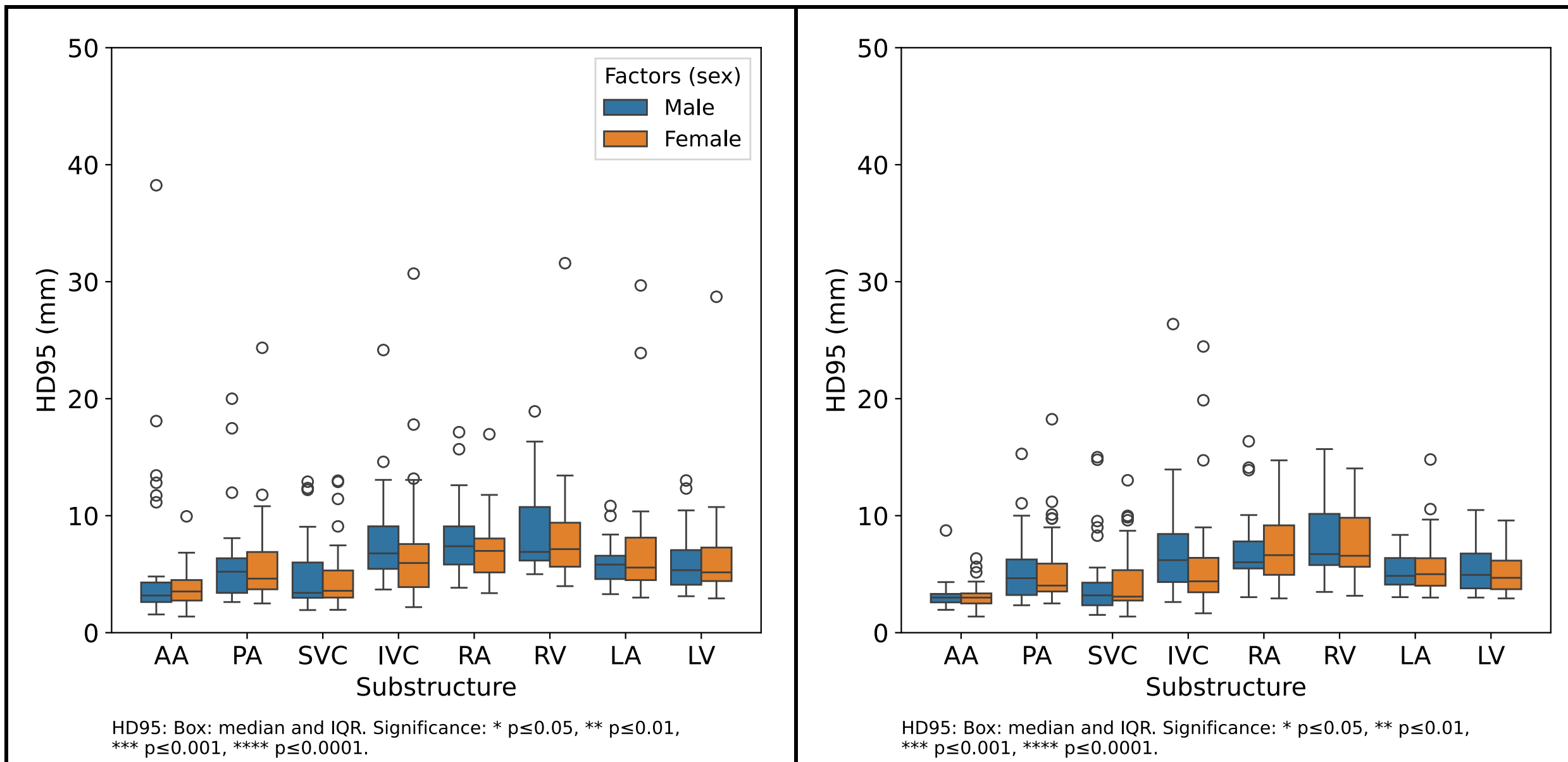

(a) SMIT-Balanced

(b) SMIT-Oracle

(c) nnU-Net-Balanced

(d) TotalSegmentator

Supplementary Figure S7: Robustness to biological sex for scans from cohort I. Box plots show HD95 (mm).

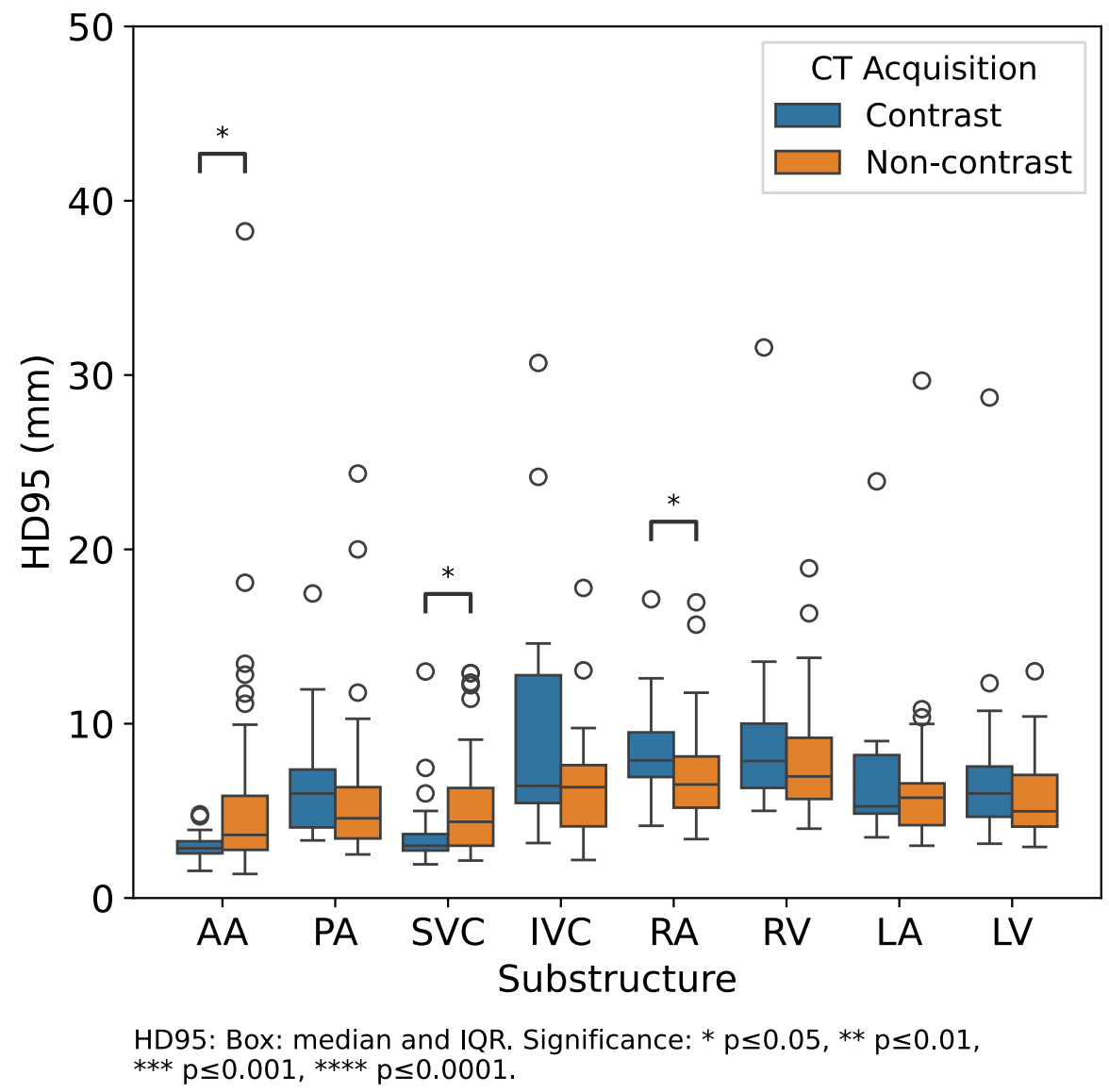

(a) SMIT-Balanced

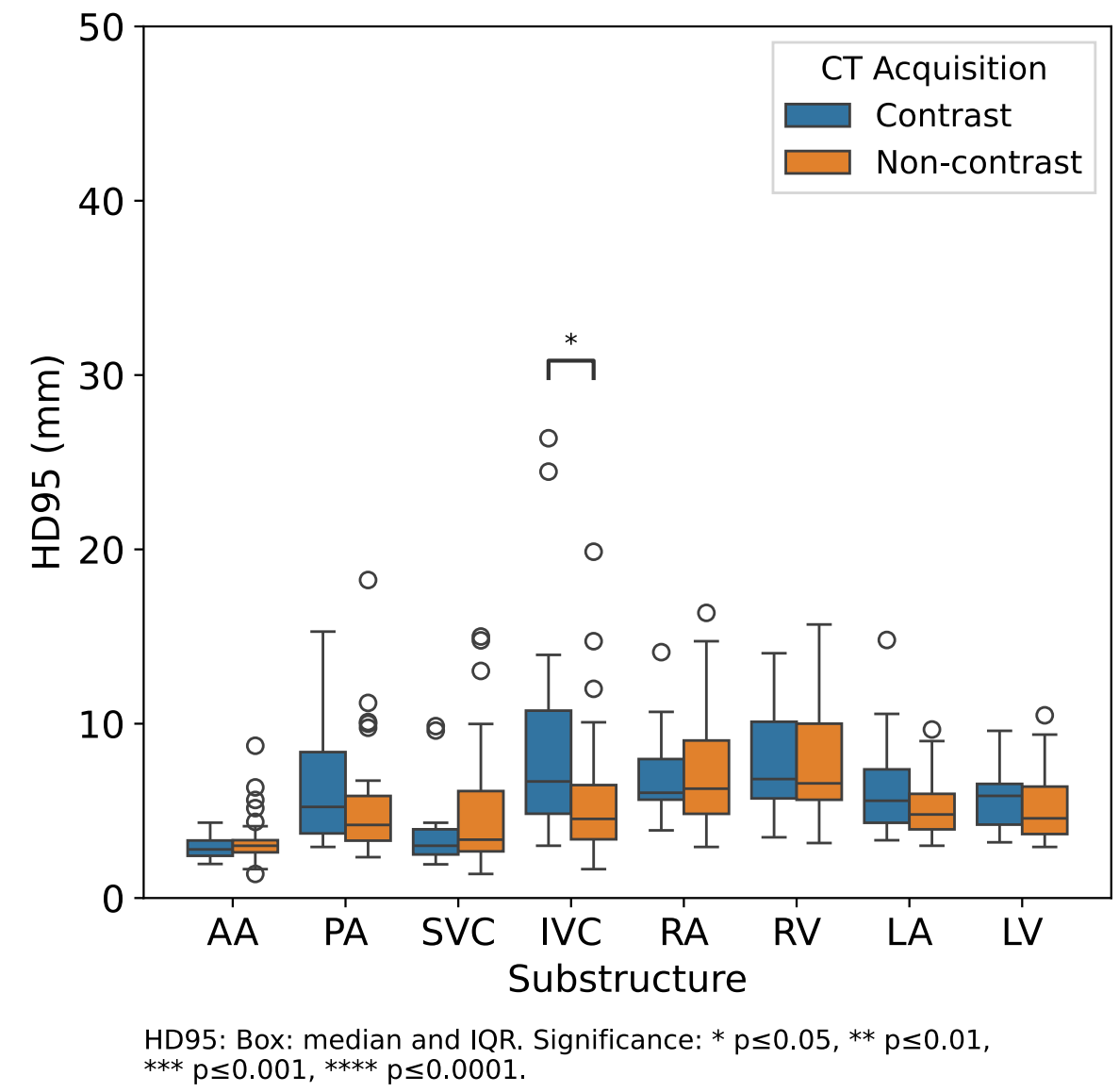

(b) SMIT-Oracle

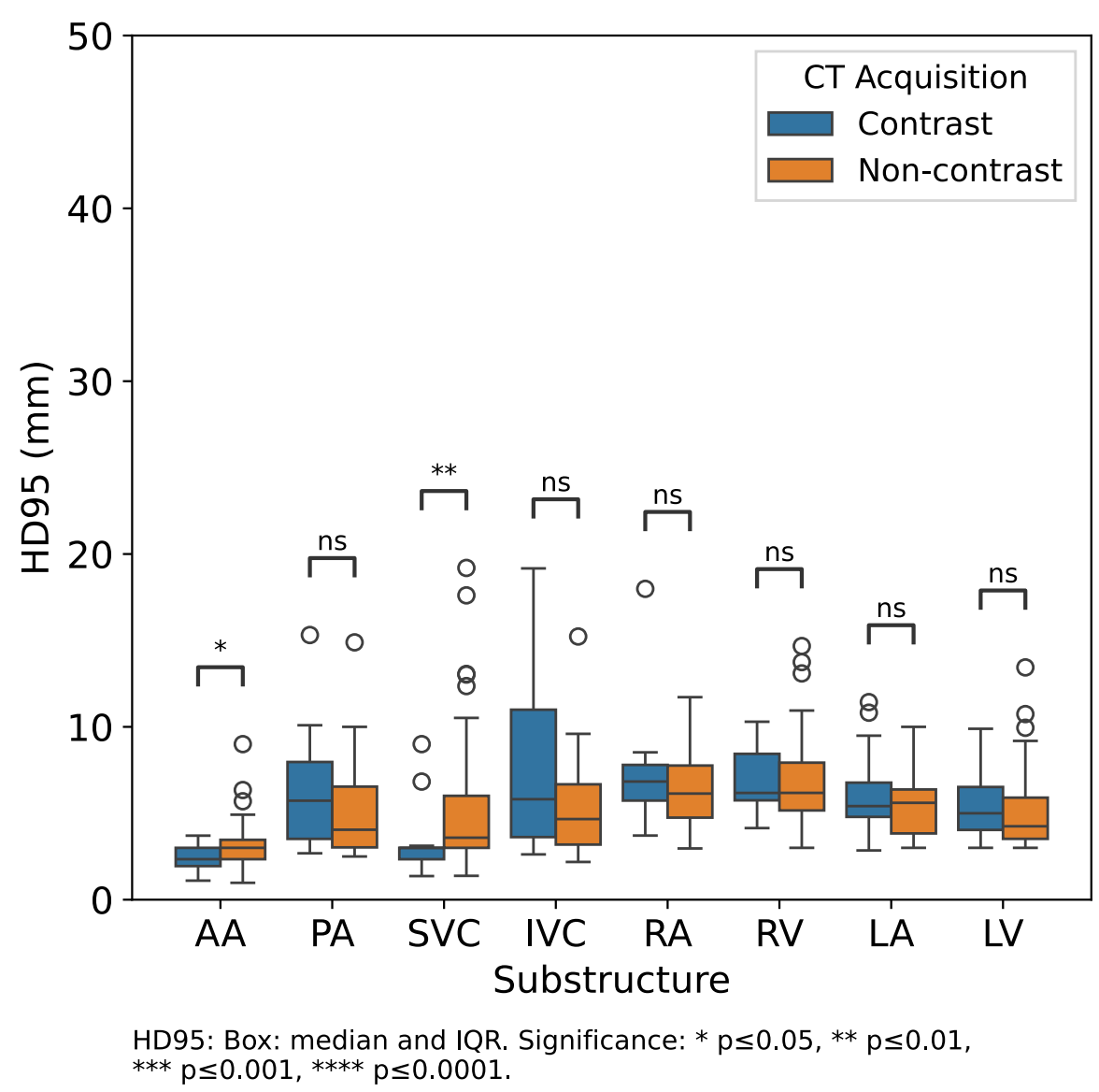

(c) nnU-Net-Balanced

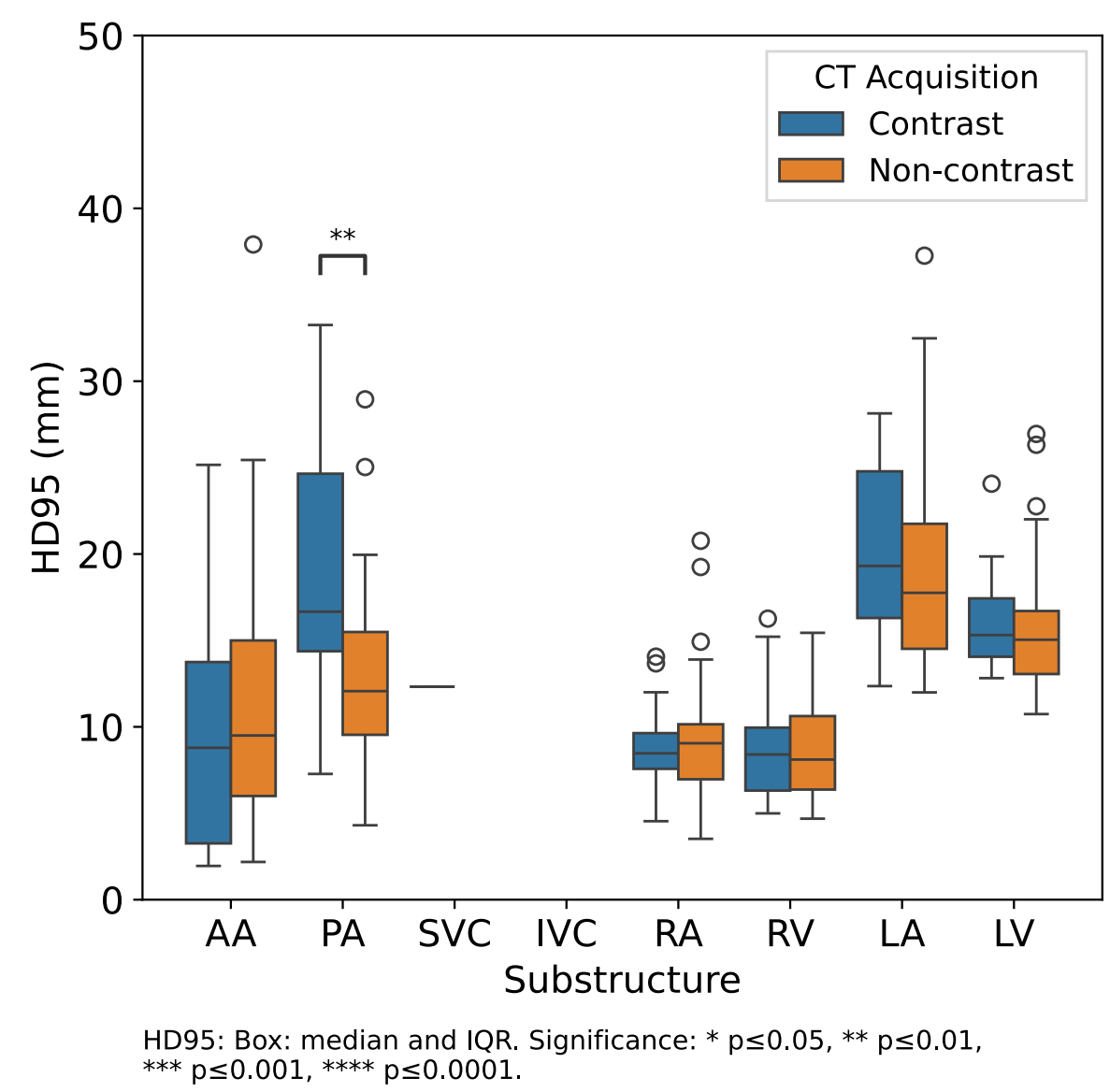

(d) TotalSegmentator

Supplementary Figure S8: Robustness to imaging contrast for scans from cohort I. Box plots show HD95 (mm).

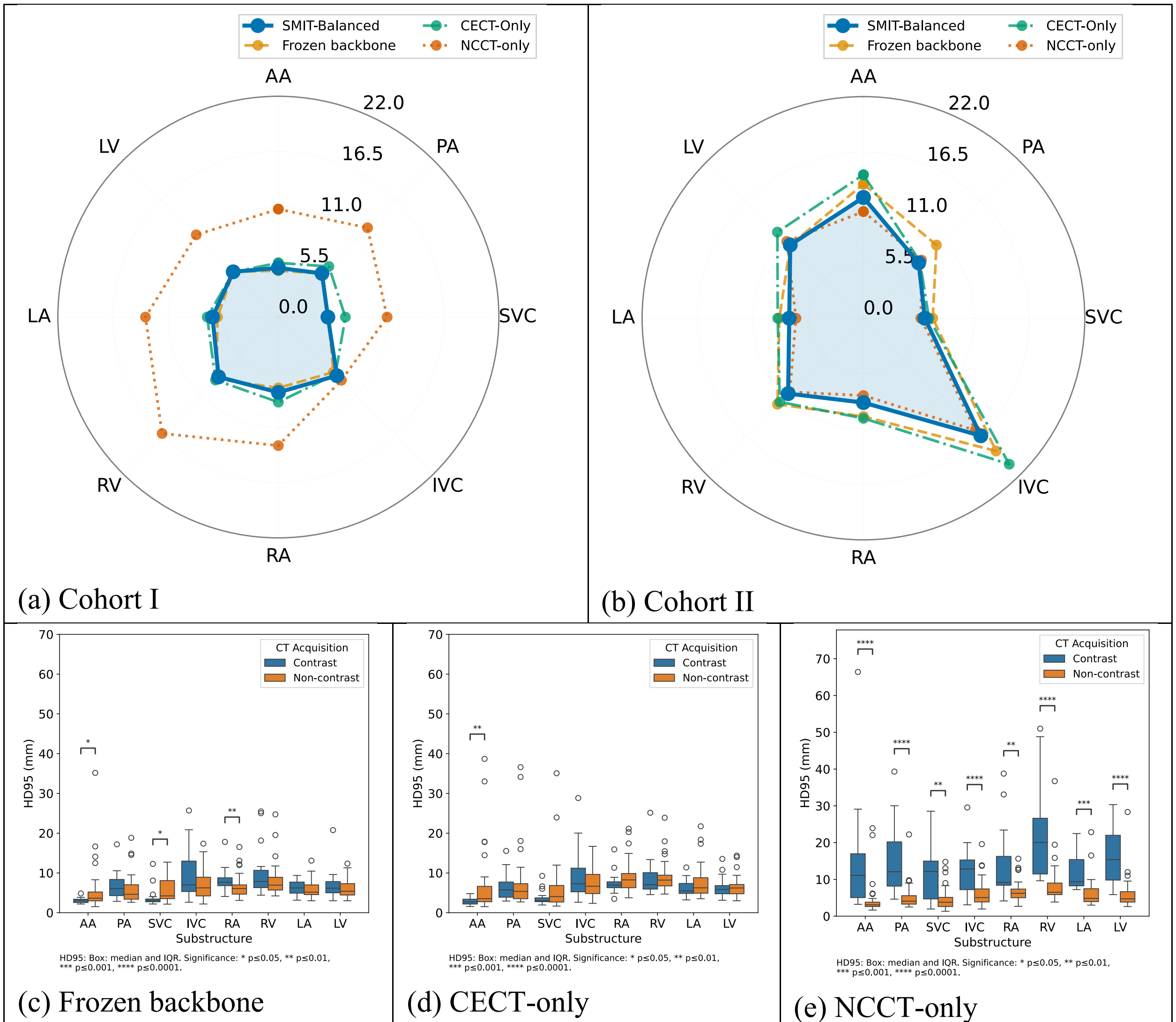

Supplementary Figure S9: Cardiac substructure segmentation performance across SMIT variants. (a,b) Radar plots show HD95 (mm) for each substructure in Cohorts I and II. (c-e) Box plots compare contrast vs. non-contrast CT for additional configurations on Cohort I.

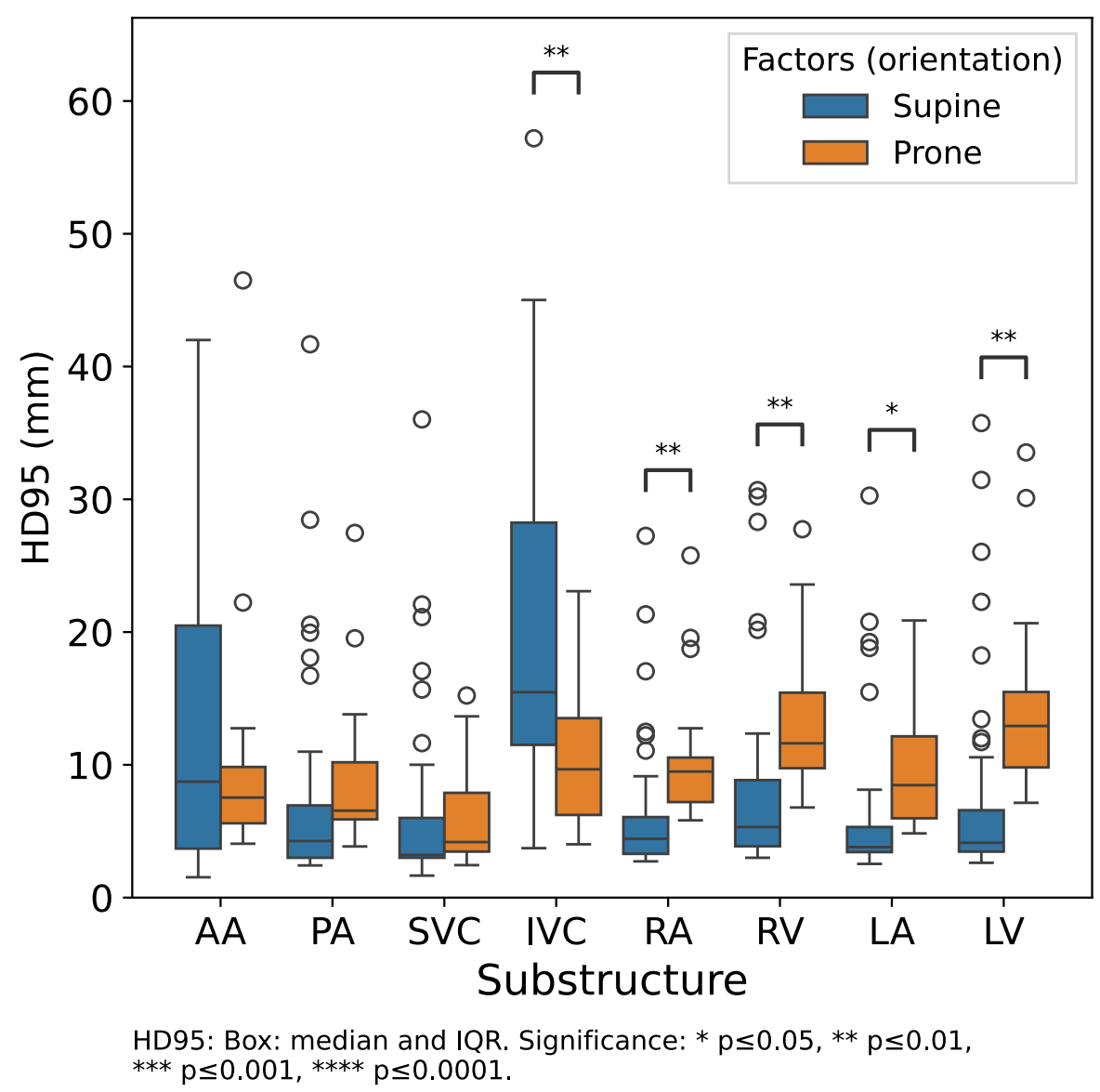

(a) SMIT-Balanced

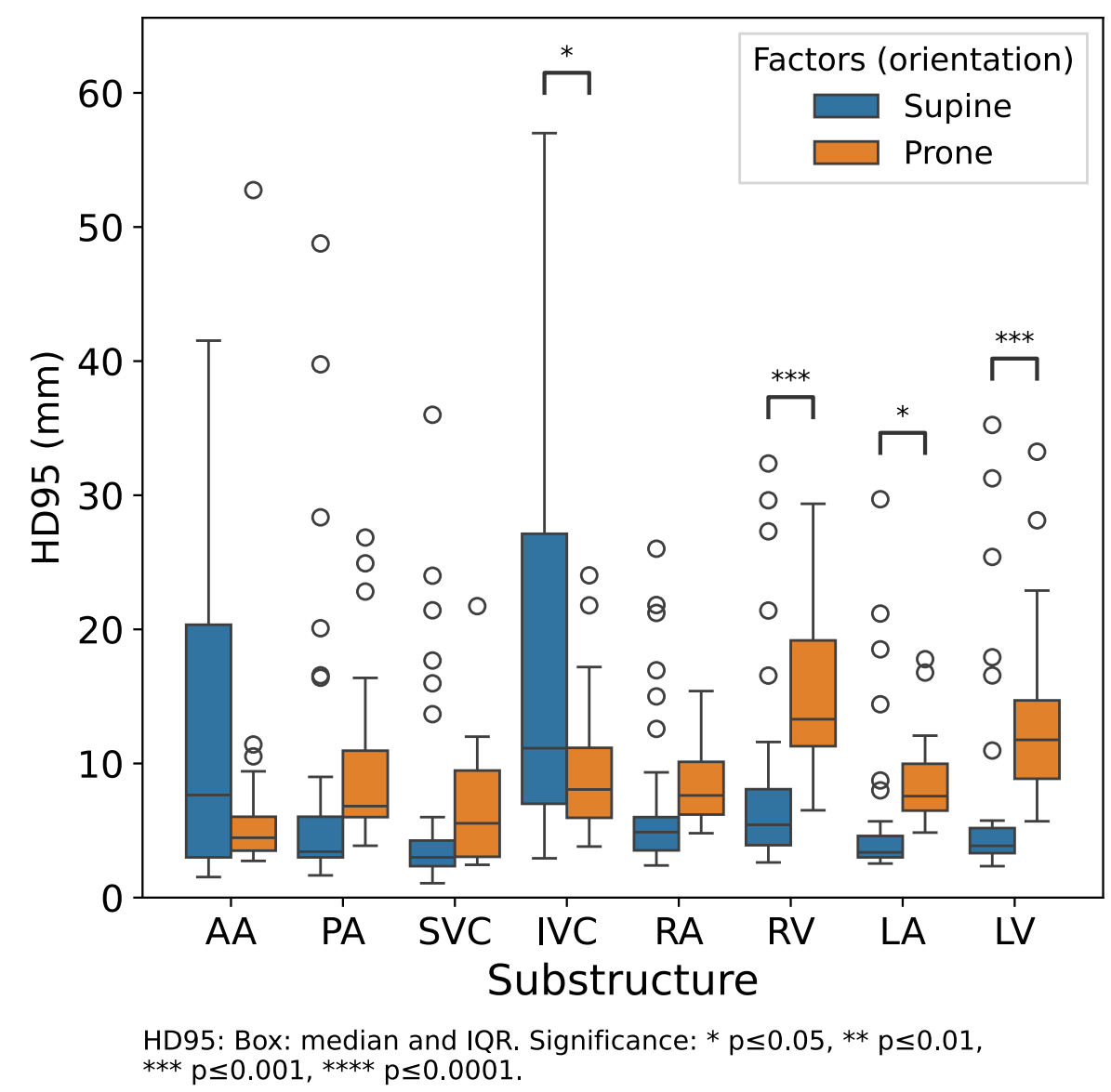

(b) SMIT-Oracle

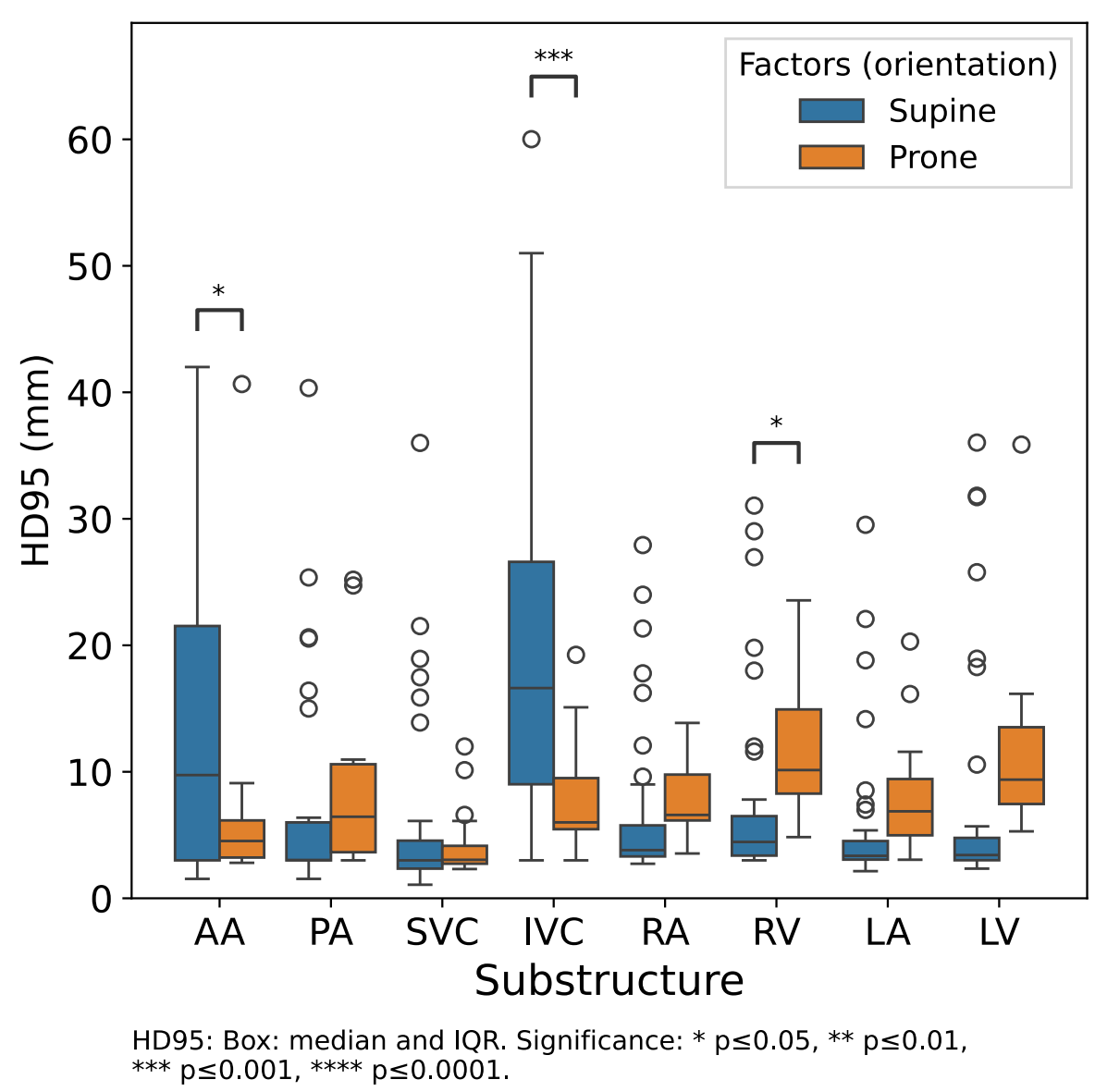

(c) nnU-Net-Balanced

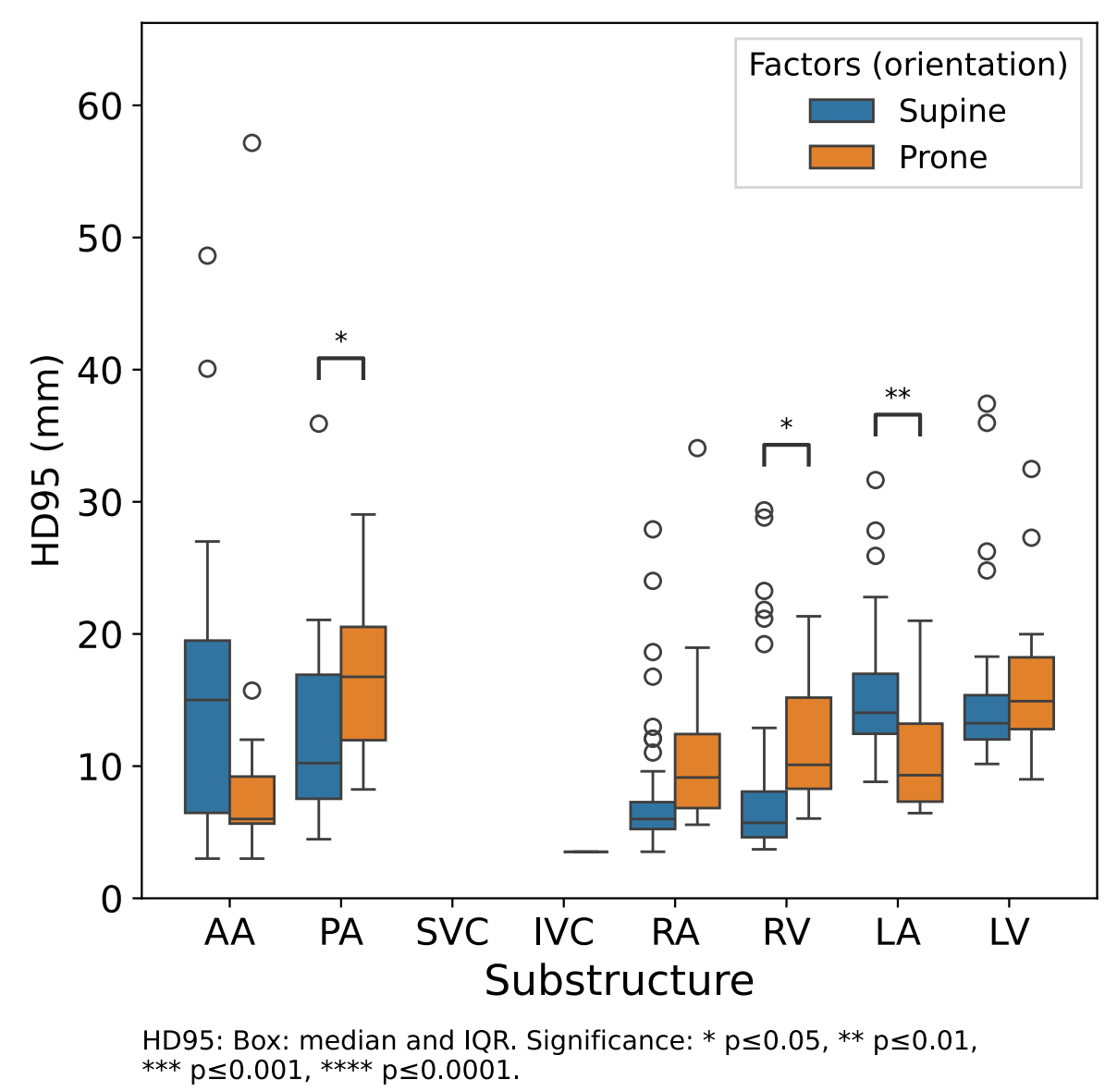

(d) TotalSegmentator

Supplementary Figure S10: Robustness to patient positioning for scans from cohort II. Box plots show HD95 (mm).

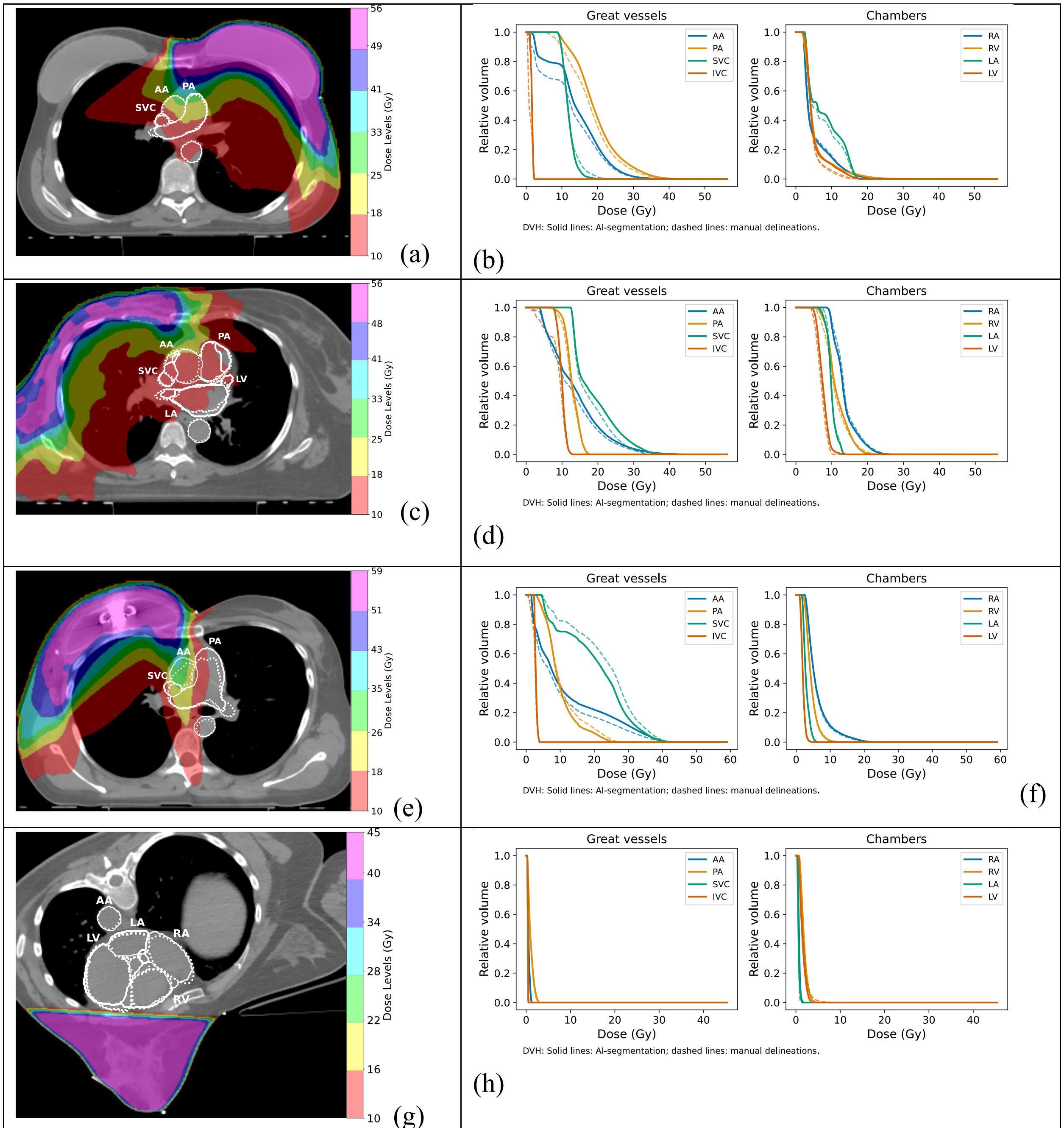

Supplementary Figure S11: Dose-volume metric comparison. Radiation dose distributions (a, c, e, g) and corresponding dose–volume histograms (DVHs) (b, d, f, h) for SMIT-Balanced model on three representative scans from Cohort II. Solid lines indicate model segmentations; dotted lines indicate manual delineations.

Supplementary Table S1: Dose-volume metric comparison between manual delineations and auto-segmentations generated by the Balanced configuration on Cohort I. Mean and standard deviation values are reported. $V_{40}$ denotes the percentage of volume receiving at least 40 Gy.

| | Metric (Gy) | Manual delineation | Balanced model | p-value |
|---|---|---|---|---|
| Aorta | $D_{Max}$ | 67.6 ± 6.7 | 67.7 ± 6.8 | 0.9 |
| Pulmonary artery | $V_{40}$ | 66.1 ± 21.4 | 65.5 ± 21.1 | 0.8 |
| Superior vena cava | $D_{Max}$ | 64.2 ± 11.4 | 64.3 ± 11.6 | 0.9 |
| Inferior vena cava | $D_{Max}$ | 8.8 ± 15.3 | 8.1 ± 14.3 | 0.7 |
| Right atrium | $D_{Mean}$ | 15.1 ± 14.1 | 15.3 ± 14.7 | 0.9 |
| Right ventricle | $D_{Mean}$ | 7.9 ± 7.5 | 7.9 ± 7.3 | 0.9 |
| Left atrium | $D_{Mean}$ | 27.6 ± 14.4 | 26.6 ± 14.1 | 0.6 |
| Left ventricle | $D_{Mean}$ | 7.4 ± 7.3 | 7.2 ± 7.1 | 0.8 |

Supplementary Table S2: Dose-volume metric comparison between manual delineations and auto-segmentations generated by the Balanced configuration on Cohort II. Mean and standard deviation values are reported. $V_{40}$ denotes the percentage of volume receiving at least 40 Gy.

| | Metric (Gy) | Manual delineation | Balanced model | p-value |
|---|---|---|---|---|
| Aorta | $D_{Max}$ | 8.2 ± 15.4 | 6.9 ± 13.2 | 0.7 |
| Pulmonary artery | $V_{40}$ | 0.1 ± 1.2 | 0.0 ± 0.0 | 0.6 |
| Superior vena cava | $D_{Max}$ | 4.6 ± 10.4 | 4.0 ± 9.4 | 0.9 |
| Inferior vena cava | $D_{Max}$ | 1.8 ± 5.7 | 1.1 ± 2.4 | 0.9 |
| Right atrium | $D_{Mean}$ | 1.3 ± 2.4 | 1.3 ± 2.3 | 0.9 |
| Right ventricle | $D_{Mean}$ | 1.3 ± 2.3 | 1.4 ± 2.3 | 0.8 |
| Left atrium | $D_{Mean}$ | 1.1 ± 2.5 | 0.9 ± 1.9 | 1.0 |
| Left ventricle | $D_{Mean}$ | 1.3 ± 2.0 | 1.3 ± 2.0 | 1.0 |

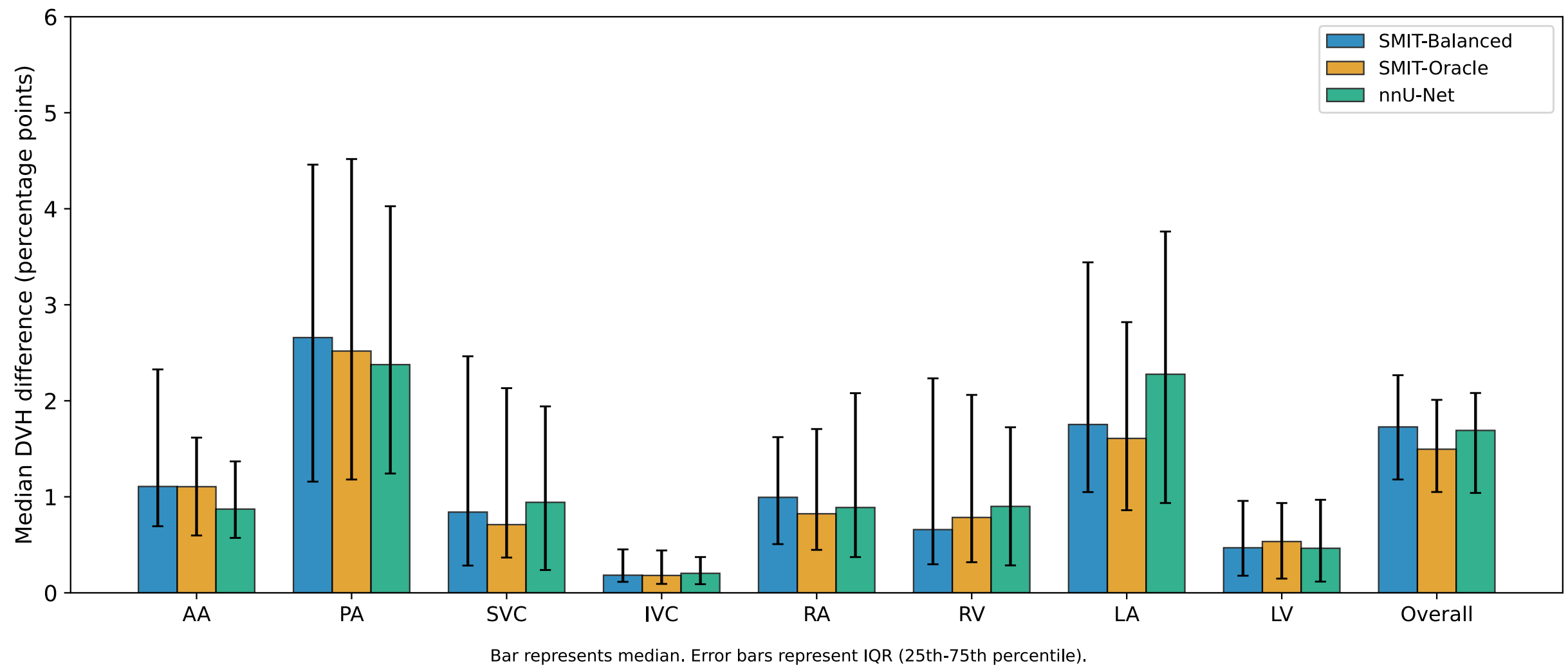

(a)

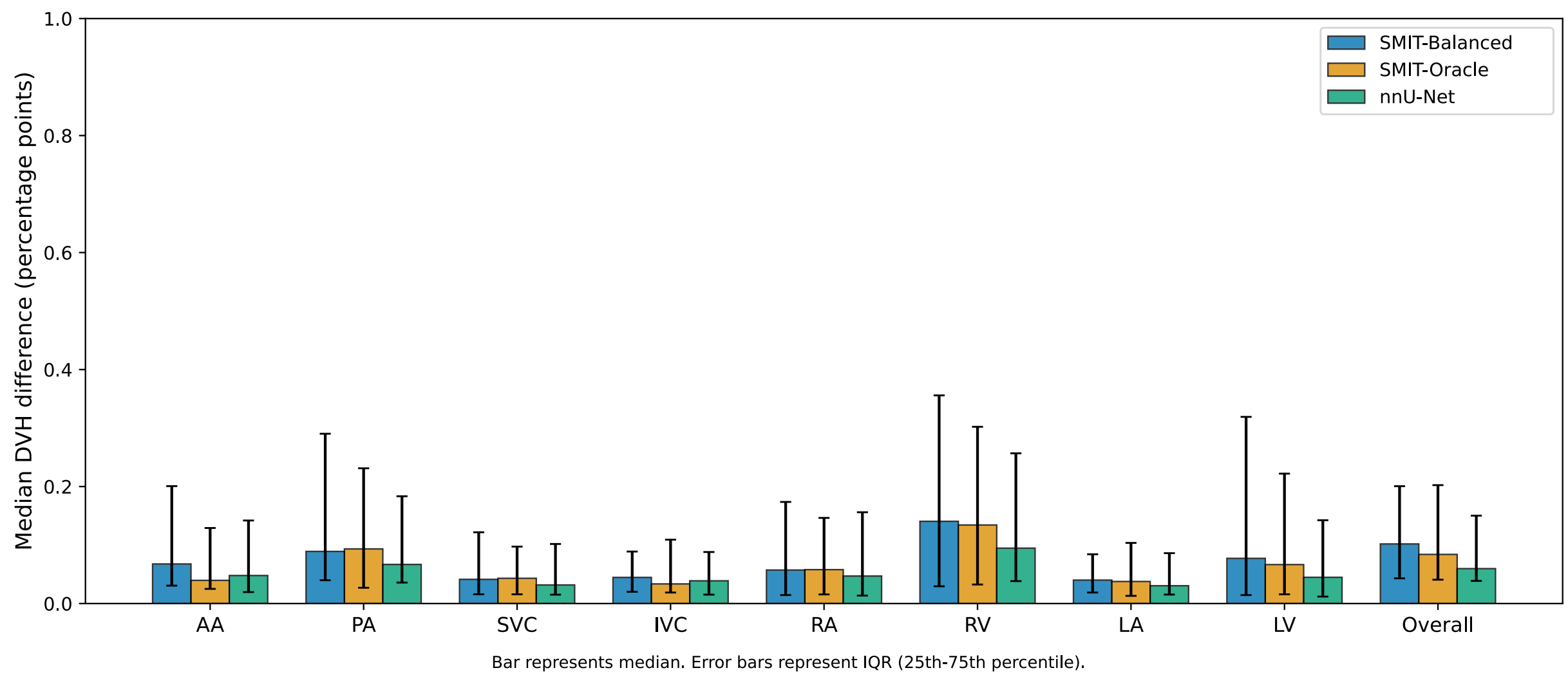

(b)

Supplementary Figure S12: DVH difference across model configurations for cohort I (a) and cohort II (b).

**Supplementary Material A**

SMIT (self-distilled masked image transformer) consists of three self-supervised learning tasks: masked voxel token prediction, masked patch token distillation, and image token distillation. Masked voxel token prediction is a dense regression task of the intensities within masked patches and is optimized by minimizing the $l_1$ norm between the predicted and ground-truth patches. Self-distillation was required for the next two tasks and performed using an exponentially moving average (EMA) teacher model with identical architecture as the student. Masked patch token distillation aligns the voxel token representations learned by the student with those extracted by the teacher from unmasked sequences, optimized with a temperature-scaled cross-entropy loss. For image token distillation, the cross-entropy between global volume embeddings of the student and teacher was minimized. After pretraining, the teacher was discarded, and the student encoder was retained for fine-tuning.

**Supplementary Material B**

Before passing each input scan through the models, Hounsfield units were clipped to the range -200 to 300, then scaled between 0 to 1. Spatial resampling to standardized voxel spacing used bilinear interpolation for images and nearest-neighbor interpolation for labels. Input scans were cropped to fixed size $128^3$ sub-volumes during training. Crops were sampled such that the cardiac region was included, using the available bounding boxes derived from the reference segmentations. During inference, predictions were generated using the same crop size with appropriate overlap and recombination.

All models were trained with the Adam optimizer using an equally weighted combination of soft Dice loss and cross-entropy loss, summed across all output channels. Data augmentations included:

- random flipping along axial, sagittal, and coronal planes (probability = 0.5 each),
- random 90° rotations (probability = 0.5, up to 270°)
- random intensity scaling (±10%, probability = 0.25), and
- random intensity shifting (±10%, probability = 0.25).

These augmentations were applied sequentially on-the-fly during training to increase effective training set diversity without requiring additional storage. The cosine learning-rate schedule with warmup of 50 epochs was used to ensure that the learning rate was smoothly annealed from its initial value to a minimum value following a cosine function over the course of training.

**Supplementary Material C**

nnU-Net normalized the images based on the dataset statistics between 0 and 1, and employed its default extensive augmentation strategy, comprising of geometric and intensity transforms applied with specific probabilities:

- Random rotations (probability = 0.20) with rotation angles automatically determined based on dataset properties,
- random scaling (probability = 0.20) with scale factors ranging from 0.7 to 1.4,
- random axis-aligned mirroring (flipping) along anatomically appropriate axes,
- gaussian noise addition (probability = 0.20, variance 0 – 0.1),
- gaussian blurring (probability = 0.20, $\sigma$ of 0.5 – 1.0),
- brightness adjustment (probability = 0.15, multiplier 0.75 – 1.25),,
- contrast adjustment (probability = 0.15, range 0.75 – 1.25)
- gamma correction with intensity inversion (probability = 0.10, $\gamma$ of 0.7 – 1.5),
- gamma correction without intensity inversion (probability = 0.30, $\gamma$ of 0.7 – 1.5), and
- low-resolution simulation (probability = 0.20, down-sampling scale 0.5 – 1.0),

All augmentation parameters were determined by nnU-Net's automated configuration process based on dataset characteristics. This aggressive augmentation strategy is a key component of nnU-Net's strong generalization performance but requires substantially longer training time compared to SMIT (250,000 vs 32,000 iterations) as it is trained from scratch and customized to the dataset in question.